\documentclass[iop]{emulateapj}
\usepackage{graphicx}
\usepackage{amssymb}
\usepackage{epsfig}
\usepackage{epstopdf}
\usepackage{gensymb}
\usepackage{amsmath}
\usepackage{threeparttable}

\usepackage{txfonts}
\usepackage{longtable}

\usepackage{natbib}
\usepackage{url}

\newcommand{\etal}{{\em et al.~}}
\newcommand{\textsec}{{\textsection}}

\shorttitle{Mass Substructure in Abell 3128}
\shortauthors{McCleary et al.}
\begin{document}

\submitted{Accepted for publication in ApJ.}
\title{Mass Substructure in Abell 3128}
\author{J. McCleary, I. dell'Antonio, \& P. Huwe} \affil{Department of Physics, Brown University, Box 1843, Providence, RI 02912, USA}
\email{Jacqueline\_McCleary@brown.edu}

\begin{abstract}

We perform a detailed 2-dimensional weak gravitational lensing analysis of the nearby ($z=0.058$) galaxy cluster Abell 3128 using deep $ugrz$ imaging from the Dark Energy Camera (DECam).  We have designed a pipeline to remove instrumental artifacts from DECam images and stack multiple dithered observations without inducing a spurious ellipticity signal.  We develop a new technique to characterize the spatial variation of the PSF which enables us to circularize the field to better than 0.5\% and thereby extract the intrinsic galaxy ellipticities.  By fitting photometric redshifts to sources in the observation, we are able to select a sample of background galaxies for weak lensing analysis free from low-redshift contaminants. Photometric redshifts are also used to select a high-redshift galaxy subsample, with which we successfully isolate the signal from an interloping $z = 0.44$ cluster. We estimate the total mass of Abell 3128 by fitting the tangential ellipticity  of background galaxies with the weak lensing shear profile of an NFW halo, and also perform NFW fits to substructures detected in the 2-D mass maps of the cluster. This study yields one of the highest resolution mass maps of a low-$z$ cluster to date, and is the first step in a larger effort to characterize the redshift evolution of mass substructures in clusters. 
\end{abstract}


\section{INTRODUCTION}

Cosmological perturbation theory provides a framework in which cold dark matter organizes itself hierarchically, first collapsing into small structures which can overcome cosmological expansion and then continuing to merge into increasingly large halos. Because small collapsed objects often survive accretion onto a larger system to become sub-halos of their host, the hierarchical structure formation paradigm predicts that dark matter halos should be rich in mass substructures (\citealp{2011ApJ...740..102K, 2012MNRAS.425.2169G}).

The amount of mass substructure that we observe should increase with redshift, particularly for cluster-sized halos: mergers of galaxy- and group-size halos are more common in the early Universe, and the resulting clusters have long dynamical relaxation times (\citealp{2004MNRAS.355..819G}). At higher redshifts, then, we expect to observe an increasingly high fraction of the total cluster mass locked up in sub-regions of localized mass enhancement. An observational study of cluster substructure and its cosmic evolution would probe the assembly history of cluster-sized halos and test the CDM paradigm on sub-megaparsec scales. Characterizing substructure in clusters also has important implications for understanding the role of the mass environment in the evolution of member galaxies. Correlating sub-halo locations and, e.g., star formation rates would reveal the effect of local mass environment (distinct from the larger-scale mass distribution) on galaxy properties.  

Studies of cluster substructure are already underway. Most notably, X-ray data have been used to obtain cluster mass functions through the proxies of cluster gas emissivity and temperature. However, these proxies are related to mass by scaling relations that rely on assumptions like the hydrostatic equilibrium of the intra-cluster medium. Accretion-induced heating of cluster gas, as well as merger-induced bulk and turbulent motions, violate the assumption of hydrostatic equilibrium at the substructure level, at which the cluster is dynamically active. Moreover, the error induced by the assumption is greater in the outskirts of higher redshift clusters, where mergers are more frequent and clusters are accreting more rapidly (\citealp{2013ApJ...777..151L, 2014ApJ...782..107N}). Hence, an accurate observational study of cluster substructure requires an analysis technique insensitive to the dynamical state of a cluster.

Because of its freedom from assumptions about baryonic physics, weak gravitational lensing (WL) has become a standard tool for measuring mass concentrations in the Universe. 
Multiple observations show that individual dark matter substructures within a cluster are capable of producing their own detectable weak lensing shear (\citealp{2008PASJ...60..345O, 2008ApJ...673..163S, 2012AAS...22043510H}). However, in the weak lensing regime, the distortion of background galaxies induced by the intervening cluster is much smaller than the intrinsic uncertainty of galaxy shape measurement. To overcome this challenge, obtaining the angular resolution needed to identify cluster substructures, requires large numbers of resolved background objects. We achieve this in our observations by taking deep, wide-field images using the Dark Energy Camera (DECam) on the Blanco 4--m telescope at Cerro Tololo International Observatory.

In this paper, we describe the analysis pipeline we developed to make weak lensing measurements on DECam data and present the results of our first substructure study on a nearby ($z=0.06$) cluster, Abell 3128,  which is one of the highest resolution mass maps of a $z < 0.1$ cluster to date (see also \citealp{2014ApJ...784...90O}). We chose to begin our investigation of cluster substructure with Abell 3128 for several reasons.  First, weak lensing analysis requires clusters to be massive enough that their (obviously less massive) substructures can be detected, and A3128 is one of the most massive clusters in the Local Volume without a published convergence map. Second, the large projected size of low-$z$ clusters (and their substructure) means that their distortions will be coherent over a large swath of the sky. This greatly increases the number of background galaxies from which to measure the WL signal, and makes substructure easier to detect at low redshifts. Finally, Abell 3128 has a complex morphology that has been well studied at radio, optical and X-ray wavelengths (\citealp{2002AJ....123.1216R, 2007A&A...474..707W}), enabling the comparison of our WL substructure analysis to other techniques. 

 Our observations of Abell 3128 and reduction method are described in \textsection 3 and \textsection 4, respectively. In \textsection 4 we also discuss our weak lensing analysis and substructure identification schemes, whose results we present in \textsection 5. In \textsection 6 we explore the significance of our findings. Finally, we summarize our conclusions in \textsection 7. 

\section{WEAK GRAVITATIONAL LENSING\label{sec:WLformalism}}

A thin gravitational lens deflects and distortsthe images of background sources. It is customary to describe the mass distribution of the lens in terms of its surface mass density $\Sigma$ or its dimensionless {\it convergence}, 
\begin{equation}
\kappa = \frac{\Sigma}{\Sigma_{crit}},
\end{equation}
where the critical surface mass density $\Sigma_{crit}$ of the lens is defined as 
\begin{equation}
\Sigma_{crit} =\frac{c^2}{4\pi G}\frac{D_s}{D_l D_{ls}}\label{sigmacrit}.
\end{equation}
The quantities $D_s$, $D_l$ and $D_{ls}$ are the angular diameter distances to the background source (galaxy), the lensing cluster, and between the source (galaxy) and lensing cluster, respectively.

Lenses with $\kappa \ge 1$ produce the dramatic arcs, Einstein rings, and multiple images of strong gravitational lensing (\citealp{2010ApJ...723.1678C}). Areas of $\kappa < 1$ define the weak lensing (WL) regime, in which the distortion of background images produced by the lensing cluster is much smaller than the background images themselves. Because $\kappa \ge 1$ typically only in the dense core of a cluster, to study substructure over a cluster's entire virial region requires WL analyses. 

The convergence $\kappa$ produces the isotropic magnification of a background source, and as such cannot be measured directly from an image without some prior knowledge of the source size. In the case of weak lensing, however, convergence can be recovered from the {\it shear} $\gamma$ of background source images caused by the tidal forces of the lens' gravitational field. In particular, convergence is related to shear through the application of the inverse 2-D Laplacian to the 2-D gravitational potential of the lens, an integral transform first derived in~\cite{1993ApJ...404..441K} and~\cite{1995ApJ...449..460K}, with variants published by~\cite{1994ApJ...437...56F} and~\cite{1997AJ....114...14F}. 



In the WL regime ($\kappa <<1, \gamma<<1$), the background galaxy images experience a curl-free stretching in the direction tangential to the line-of-sight from the lens. Now,  galaxies are elliptical objects with a measured ellipticity $e = 1-(b/a)$ tilted at some position angle $\theta$ with respect to the image axes.  Galaxy shapes are frequently decomposed into ellipticity moments: $e_1= e \cos(2 \theta)$ is the projection of a galaxy's ellipse onto the image $x$ and $y$ axes, and $e_2= e\sin(2 \theta)$ is its projection onto the lines $y=x$ and $y=-x$. With this in mind, the the shear induced by a lens can be written in terms of the galaxies' shapes as
\begin{equation}
\gamma \rightarrow e_{\tan} = -(e_1{}\cos(2\phi) + e_2{}\sin(2\phi)) \label{etan}.
\end{equation}
In this so-called {\it tangential ellipticity}, $\phi$ is the angle from a fiducial lens center to the galaxy measured counterclockwise from north. In other words, the factors $e1$ and $e2$ characterize a galaxy's shape and position angle relative to the image axes, and the factors of $\sin(2\phi)$ and $\cos(2\phi)$ rotate the galaxy's $e_1$ and $e_2$ into the line tangent to the radial extending from the chosen lens center. In the absence of a gravitational lens, the average tangential ellipticity $\langle e_{\tan}\rangle$ should vanish when considered over many background galaxies with no intrinsic alignment. Hence, the $\langle e_{\tan}\rangle$ measured at a point in the image is an unbiased estimator of the WL shear. By measuring the a systematic deviation from zero average ellipticity with a sample of galaxies widely separated from each other in redshift space, we may reconstruct a cluster's convergence. If in addition the redshifts of the cluster and background galaxies are known, the cluster's surface mass density may be recovered.  For a comprehensive treatment, see reviews by \cite{2001PhR...340..291B} and \cite{2002LNP...608...55W}.

 In this study, we identify and characterize mass peaks using the {\it aperture mass statistic} $M_{ap}$ first introduced by \cite{1996MNRAS.283..837S}. Measured some angular distance $\theta$ away from the cluster center, $M_{ap}$ is given by the convolution of the convergence $\kappa$ with an {\it aperture mass filter} $U(|\theta_0 - \theta|)$:
 \begin{equation}
M_{ap}(\theta_0) = \frac{1}{n}\int d^2\theta\kappa(\theta)U(|\theta_0 - \theta|)\label{Map1}.
\end{equation}
The aperture mass filter $U(|\theta_0 - \theta])$ smooths the convergence over some characteristic aperture $\theta_0$. By design, the $M_{ap}$ is a local measurement involving only the shear from galaxies within an angle $\theta_0$ of the center at position$\theta$; the filter is ``compensated" so that its first order moment vanishes on scales larger than the aperture size.

If the aperture mass filter $U(|\theta_0 - \theta])$ in Equation~\ref{Map1} is transformed as  
\begin{equation}
Q(|\theta_0 - \theta|) = \frac{2}{\theta^2}\int_0^{\theta}\theta' U(|\theta_0 - \theta'|) d\theta' - U(\theta),
\end{equation}
we can replace $\kappa$ in the aperture mass statistic with the tangential ellipticity of background galaxies $e_{\tan}$. For a discrete dataset of background sources, the aperture mass thus has the form  
\begin{equation}
M_{ap}(\theta_0) = \frac{1}{n}\sum_i e^{\tan}_i(\theta)Q(|\theta_0 - \theta|),\label{aperturemassstat}
\end{equation}
where the sum is taken over all galaxies in the observation (\citealp{1996MNRAS.283..837S}). We apply Equation~\ref{aperturemassstat} to our observations to build WL convergence maps.

A variety of aperture mass filters exist which can be used in Equation\ref{aperturemassstat} exist, but the one best suited to our search for substructure was introduced by \cite{2004A&A...420...75S} as part of the GaBoDS survery. The Schirmer filter was originally designed to pick out shear signal from clusters embedded in large-scale structure, and is given by
\begin{equation}
Q(x) = \frac{1}{(1+e^{a-bx} + e^{dx-c})}\frac{\tanh(x/x_c)}{\pi R_S^2(x/x_c)} \label{Schirmerfilter},
\end{equation}
where $R_S$ is the Schirmer filter size and $x=r/R_S$ is a scaled distance between the cluster center and the point in consideration (\citealp{2005A&A...442...43H}). To optimize the filter for detection of NFW shear profiles, the parameters in Equation \ref{Schirmerfilter} are tuned to $a=6, b=150, c=47, d=50$ and $x_c=0.1$. 
By tuning the size of the Schirmer filter ($R_S$) in the aperture mass statistic, we can discern both the main cluster signal and its substructures while simultaneously characterizing their respective scales: noting that the Schirmer filter weights peak sharply at a value of $x_cR_S$, the structures identified have size $\sim 0.10R_S$.  Since the Schirmer filter is not monotonically decreasing, it is difficult to assign it with a Gaussian-type FWHM. Instead, a smoothing length can be obtained by computing the radius which encompasses $50\%$ of the filter's weight. For the form of the Schirmer filter given above, this radius is $0.121R_S$, equivalent to a Gaussian with standard deviation $\sigma$ = 0.18 or a FWHM of 0.3.

We note finally that, in analogy with electromagnetism, the curl-free tangential ellipticity defined in Equation~\ref{etan} is sometimes called the E-mode WL signal. A complementary, curl-like ellipticity referred to as the B-mode signal is obtained by rotating Equation~\ref{etan} through $\pi/4$ radians:
\begin{equation}
e_{c} =e_2{}\cos(2\phi) - e_1{}\sin(2\phi)\label{ec}.
\end{equation}
Since gravitational lensing creates no B-mode signal, the replacement of $e_{\tan}$ with $e_c$ is frequently used as a statistical control.

\section{OBSERVATIONS\label{sec:Obs}}
 
The DECam imager consists of 62 $2048 \times 4096$-pixel science CCDs (520
megapixels total) arranged in a hexagon\footnote{Note that as of the time the
  A3128 data was collected, one of the CCDs at the southern edge of the array
  (N30) was non-functional.} and captures 3 square degrees (2.2 square degrees
wide field) at  0.265''/pixel resolution~(\citealp{2008SPIE.7014E..13D, 2012SPIE.8446E..11F}). The camera's wide field of view allows us to image the entire virial region of even a low-redshift cluster in a single pointing, making it efficient for our study of cluster substructure. 

Observations of A3128 were made over eleven days from 8th-24th November 2012 in the $ugrz$Y filter set by Dara Norman and the DECam science verification team as part of that instrument's science verification program. To ensure sky coverage in the gaps between science CCDs, the telescope was dithered in a ``center + rectangle" pattern. The dithers are large enough to overlap adjacent CCDs by several hundred pixels, providing more uniform depth at the chip edges and allowing construction of a catalog covering the entire $1.5\degree \times 1.5\degree$ uniformly. The exposure time of each pointing varied by filter: 720 s in $ u$, 600 seconds in $ g$, 300 seconds in $r$ and 240 seconds each in Y and $z$. The final exposure times across the field were 10,800 seconds in $u$, 3600 seconds in $g$, 5400 seconds in $r$, 2630 seconds in Y and 2160 seconds in $z$, with at least two complete dithers in each band. The mean seeing in $r$ was {0.94''}, and after calibration of source number counts vs. magnitude against the Subaru-COSMOS catalog (\citealp{2007ApJS..172....9T}), the observations have a 50\% completeness depth of $m=24.97$ in $u$,  $m=24.76$ in $g$, $m=25.62$ in $r$ and $m=23.85$ in $z$. We note that the $Y$ and $z$ filter profiles overlap considerably, such that the narrower $Y$ essentially just covers the longer-wavelength portion of the $z$ filter. Imaging in these two filters is redundant for the purposes of our analysis, and so we make no use of the $Y$ band data beyond making a stacked image.

The CTIO+DECam system's sensitivity is greatest in $r$ band, and this filter also optimizes the balance between high background galaxy luminosity and reasonably low sky noise (both of which increase with increasing wavelength). Following the successful observing strategy of the Deep Lens Survey (\citealp{2002SPIE.4836...73W}), we observed A3128 in $r$ when the seeing FWHM reached  $<1.0$'' and in $ugz$Y otherwise. Accordingly, the $r$-band imaging has uniformly good resolution as well as a greater depth than the imaging in other bands. Shear measurements are thus made exclusively in the $r$ band, while other filters are used to provide color information for photometric redshifts (see \textsection\ref{sec:SourceSelection}).

\section{ANALYSIS\label{sec:Analysis}}
\subsection{Image Processing}
Reconstructing a two-dimensional mass maps from galaxy shapes is an involved procedure. The intrinsic galaxy ellipticities are $\sim30$ times larger than the distortions we are trying to measure, and {\it a priori} it is impossible to disentangle WL-induced shear from the shape of a single galaxy. Moreover, anisotropy in the PSF field shears incoming light from galaxies and obscures our weak lensing signal. For a camera as large as DECam, which reaches the edge of the focal plane of its telescope, this effect is substantial.  The number and size of DECam's CCDs also makes removing instrumental artifacts from observations a technical challenge. In the following section, we list the image processing steps undertaken to overcome these difficulties and measure the mass substructure of A3128. Although a community reduction pipeline now exists, we developed our own independent image processing pipeline as part of the science verification program for DECam. 

\subsubsection{CCD-Level Reduction}

The CCD-level image reduction applied to each exposure in the dataset includes the standard complement of overscan subtraction and trimming, bias subtraction, and dome flat field correction. These tasks were accomplished with the MSCRED package in IRAF\footnote{IRAF is distributed by the National Optical Astronomy Observatory}. We apply to CCD images an empirically-determined correction for the crosstalk that occurs as neighboring amplifiers are read out in parallel. A ``tree ring" pattern of concentric circles of light and dark pixels appears in all DECam object and flat field exposures; these are not an artifact of gain variations on the chips, but actually represent the physical shifting of charge between pixel wells. Tree rings in object exposures are successfully camouflaged by the flat fielding step, although flat fielding away the tree rings in object exposures is tantamount to turning an astrometric error into a photometric one.  Given the tiny amplitude of the tree rings ($\sim 0.2\%$ of pixel flux value), the error introduced is dwarfed by the $m \ge 0.03$ photometric uncertainty of the images. For more details regarding this and other CCD artifacts peculiar to DECam, see~\cite{2014JInst...9C4001P}.

To mitigate the $> 100''$ pointing errors in DECam science verification data, objects in the observation are matched against a list of reference celestial coordinates in the USNO-B catalog. We fit a linear relation, which may include a zero point shift, scale change, and axis rotation, between the observed positions and the reference coordinates on both coordinate axes. The fit is used to update the image world coordinate system so that it is registered to the reference coordinate system defined by USNO-B.


\subsubsection{PSF Correction\label{sec:PSF}} 

Because the DECam CCD array is so large, the point spread function (PSF) has significant and spatially varying contributions both from the curvature of the focal plane and the Blanco 4-m optics.  Such anisotropies in the point spread function can induce spurious shear signal, and so the accuracy of our mass maps relies on extremely precise characterization of the DECam PSF. Distortions in the PSF field of an image can be traced by systematic variations in the shapes of its stars since these are point sources and should appear perfectly round in an isotropic PSF field. Consequently, the first step in circularizing the PSF is the identification of stars in the observation. Next, polynomials are fit to the spatial variation of the following combinations of second-order intensity moments: $I_{xx} - I_{yy}$, $I_{xx} + I_{yy}$ and $I_{xy}$. Finally, the intensity moment fits are used to derive a PSF circularization kernel, which is convolved with image pixels in the stacking stage (cf.~\ref{sec:Stacking}) as in \cite{2002AJ....123..583B}.

The default procedure is to go through these steps for every CCD of every exposure in the dataset, and for most applications, this would yield a sufficiently circular PSF. However, the DECam PSF field is severely under-sampled by unsaturated stars in any single exposure; the CCD chips themselves are large, and A3128 is at high galactic latitude. Consequently, applying the standard PSF circularization technique to the DECam A3128 data only lowers the mean stellar ellipticity from 5\% to 1\%, which is still high enough to affect the WL shear signal. 
\begin{figure*}[ht]
 	\centering
	\includegraphics[width=0.4 \textwidth]{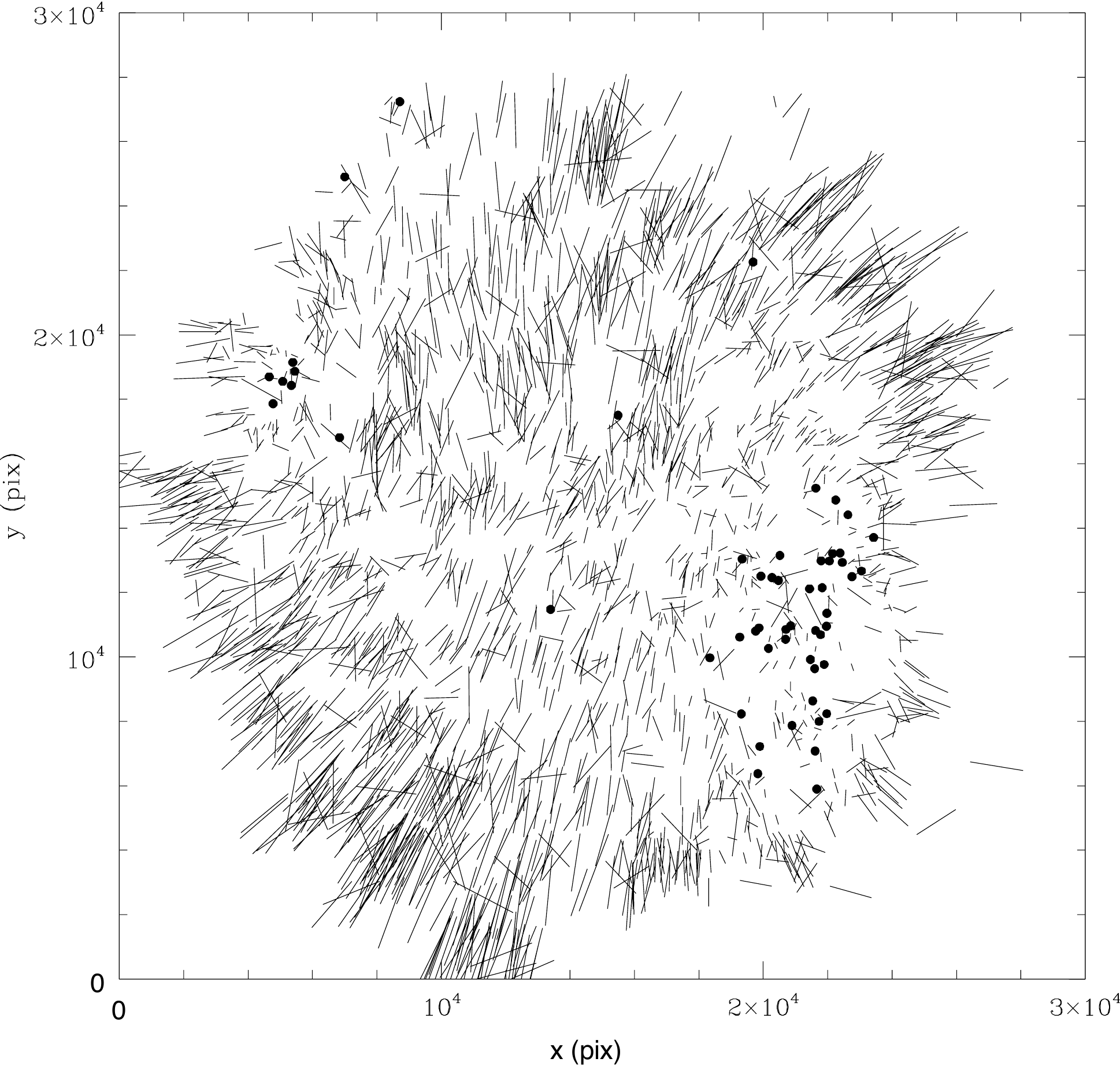}
	\includegraphics[width=0.4 \textwidth]{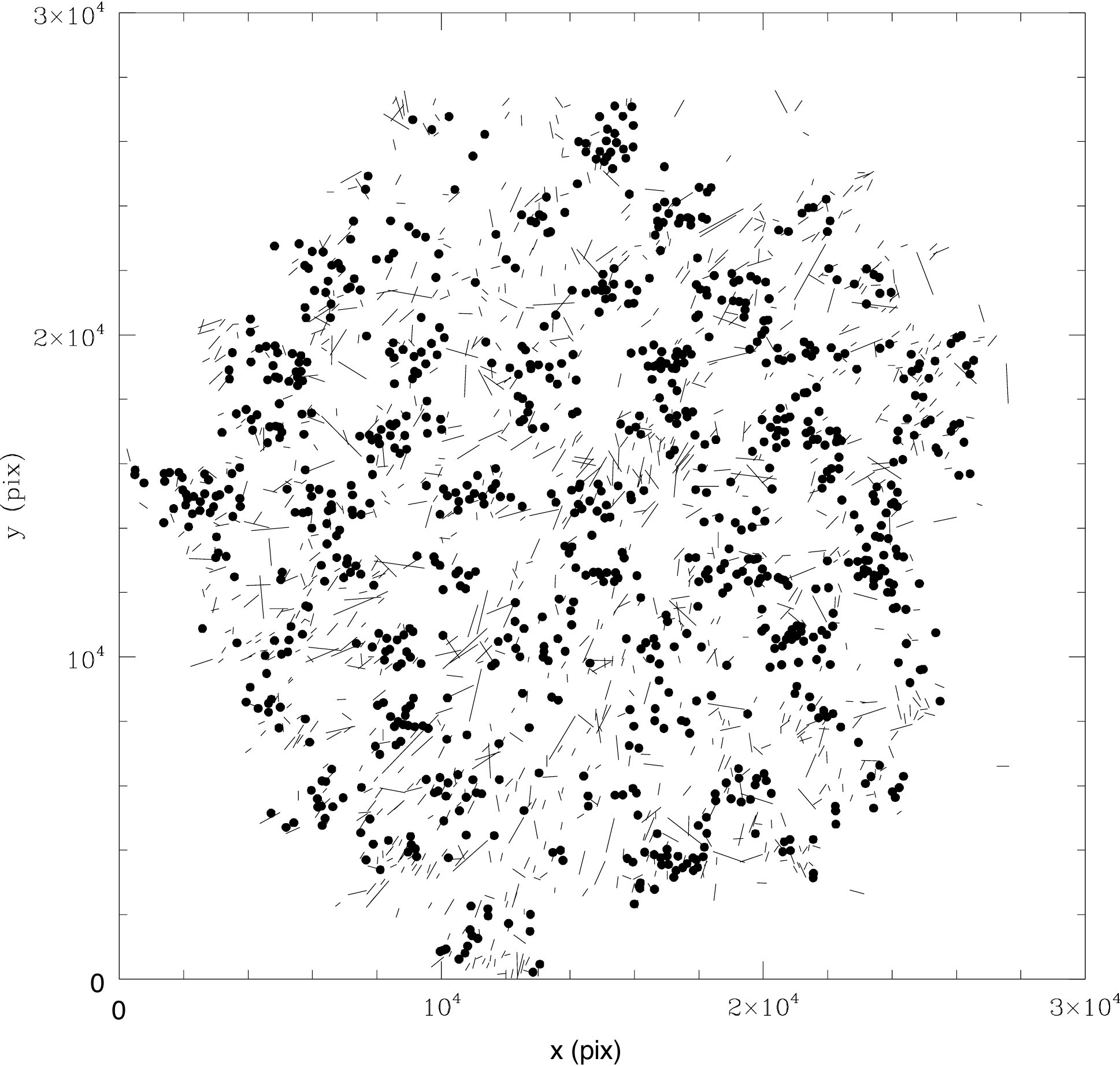}

	\caption{\footnotesize{Whisker plot showing spatial variation of the PSF
            across the Abell 3128 image. Each stick represents a star that was
            used to circularize the PSF, with length proportional to the magnitude of its measured ellipticity, and orientation equal to its position angle. Blank regions correspond to CCD chip edges, large cluster galaxies or saturated stars. {\em Left:} Stars in a Abell 3128 stack made without circularization correction. The mean segment size corresponds to stellar ellipticities $\sim$ 0.05; points represent objects with $e \le 0.008$. {\em Right:} Stack made with the multi-chip circularization correction described below. Mean ellipticity has been reduced to 0.005.}}
	\label{fig:shearsticks}
\end{figure*}

To improve the PSF modelling, we combine stellar catalogs from sequential exposures on the same CCD chip and use these ``super-catalogs" of stars to fit 2-D polynomials to the spatial variation of intensity moments. High-order polynomial terms (fourth- and fifth-order) of the intensity moment fits capture effects like focal plane curvature and tend to be stable over the course of contiguous dithers. Because the lower order terms in the intensity moment polynomials capture time-dependent effects like seeing or telescope drift, they are usually best fit using individual CCD exposure catalogs rather than super-catalogs. 

The particular grouping of exposures used to build super-catalogs is also determined empirically for each CCD. Stellar ellipticity is minimized when the higher-order terms of the $I_{xx} - I_{yy}$ polynomials are fit by super-catalogs assembled from a single dither of five exposures. Meanwhile, the higher-order terms of the $I_{xy}$ and $I_{xx} + I_{yy}$ polynomials should be fit using super-catalogs assembled from as many contiguous exposures as possible. 

For each of the 61 functional CCDs, we determine empirically both the degree of the polynomial fits and which of its terms should be obtained through super-catalogs. We piece together the final forms of fits to $I_{xx} - I_{yy}$, $I_{xx} + I_{yy}$ and $I_{xy}$ 
from whichever combination of terms ultimately yields the lowest stellar ellipticities.  This procedure allows for the circularization of stellar PSFs to better than 0.5\% (see
Figure~\ref{fig:shearsticks}). We verified that the magnitude of stars
selected does not affect the polynomial fitting by examining the residuals of
the PSF fits to stars. These show no discernible trend in the range of
magnitudes considered ($16 < m_r < 20$). 
\begin{figure*}[ht]
 	\centering
	\includegraphics[height=0.29 \textwidth]{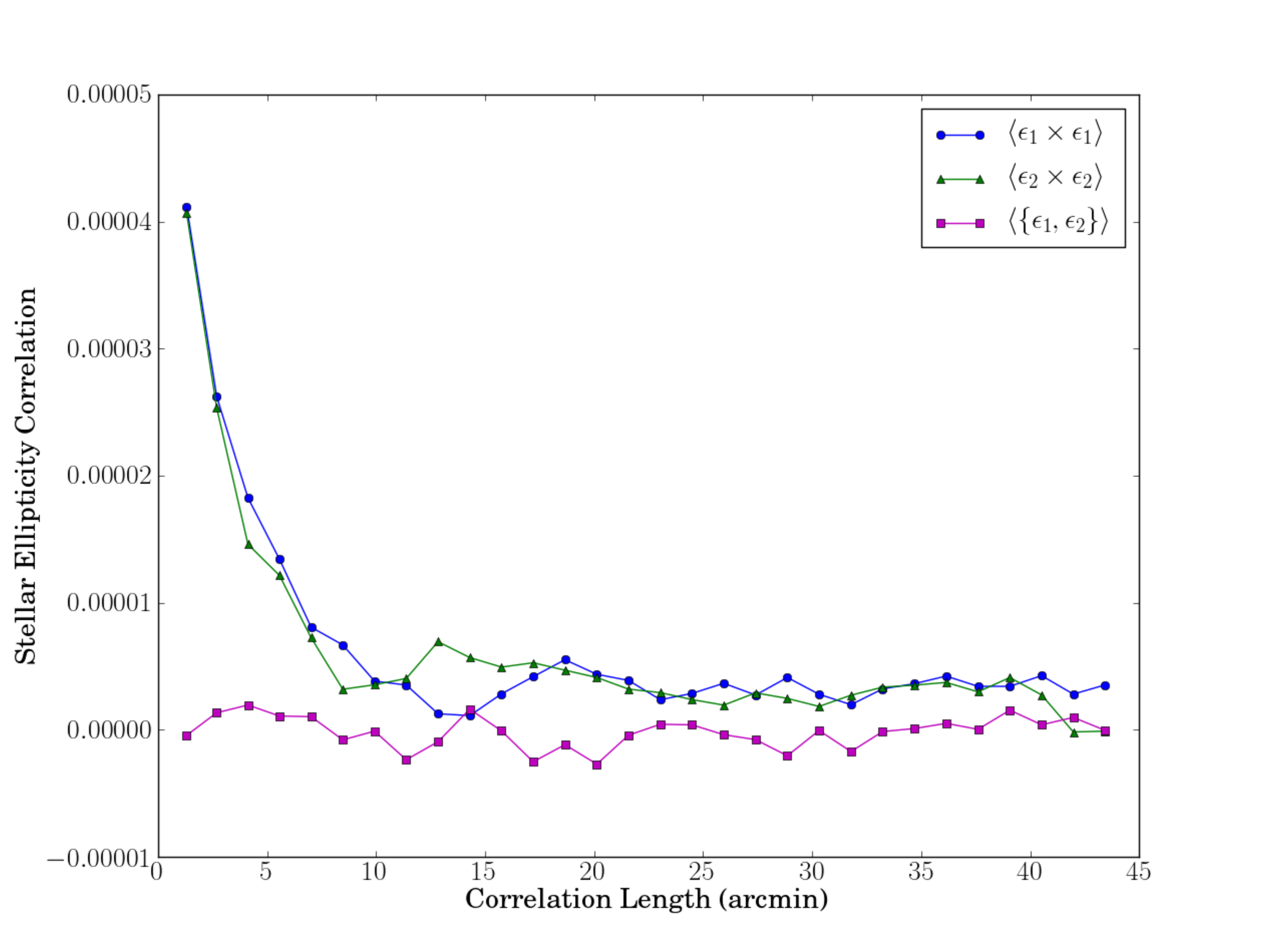}
	\includegraphics[height=0.29\textwidth]{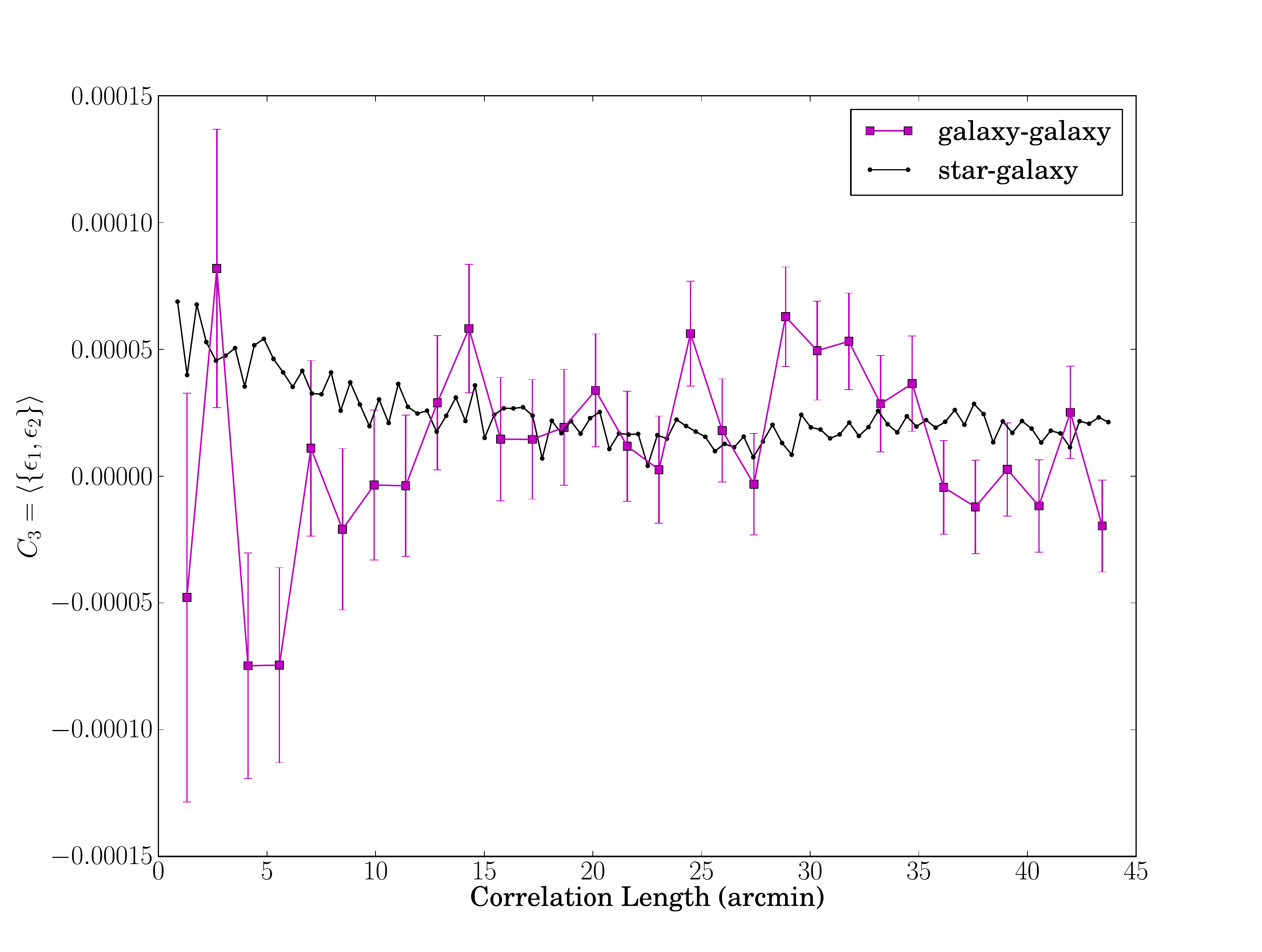}\\
	\includegraphics[height=0.29 \textwidth]{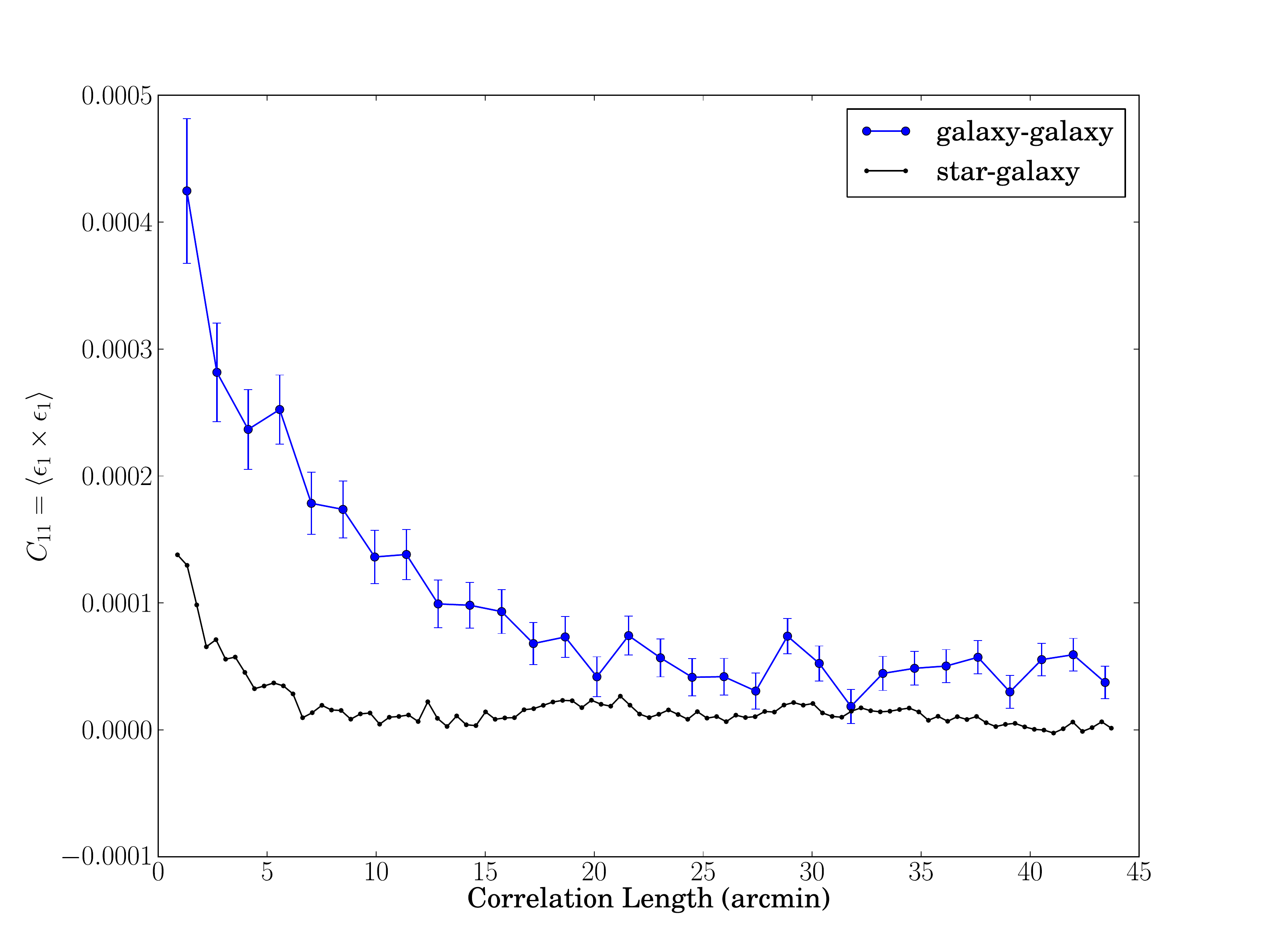}
	\includegraphics[height=0.29 \textwidth]{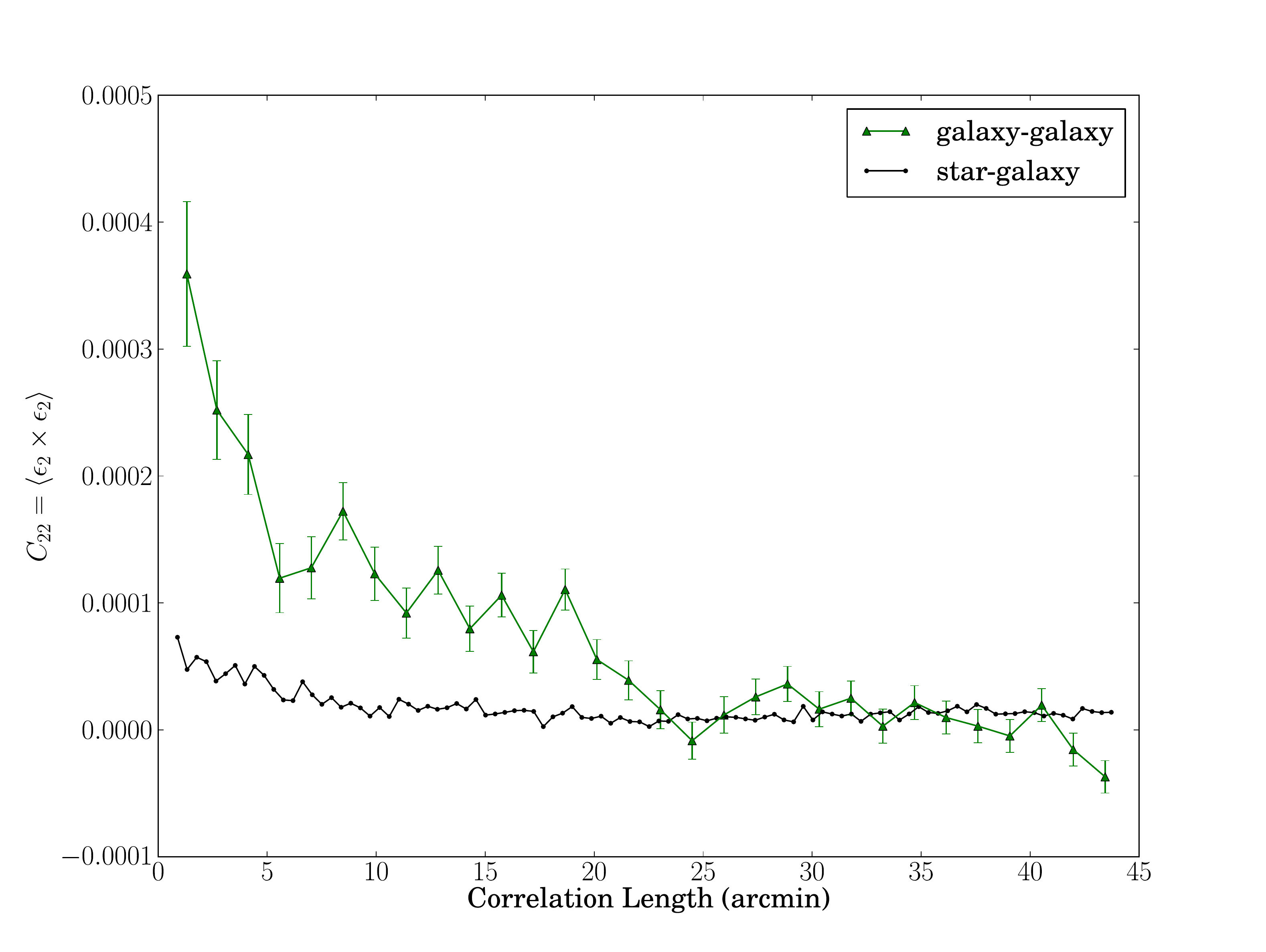}\\

	\caption{\footnotesize{Correlation functions between tangential ellipticity
            components for objects in the A3128 observation. Star-star
          auto-correlation functions are plotted at top left, and star-galaxy
          cross-correlation functions are contrasted with galaxy-galaxy
          auto-correlation functions in the other three panels. All ellipticity components are
          defined with respect to the image axes.}}
	\label{fig:correlfuncs}
\end{figure*}

 The effectiveness of this PSF
  circularization scheme may be quantified by constructing two-point shear
  correlation functions, defined as
\begin{equation}
C_{ij} = \langle e_i( {\bf r}) \times e_j( {\bf r+ \theta})\rangle,
\end{equation}
where $e_i$ is the {\it i}th ellipticity component of an object at
position  r, and brackets denote an average over all pairs within a
separation $\theta$. A third correlation function, 
\begin{equation}
C_{3} = \langle e_1( {\bf r}) \times e_2( {\bf r+
  \theta})+e_2( {\bf r}) \times e_1( {\bf r+ \theta})\rangle,
\end{equation}
should be zero and is frequently used to test for
systematic error in PSF correction schemes (\citealp{2005MNRAS.359.1277M}). Star-star auto-correlation
functions and star-galaxy cross-correlation functions are shown in
Figure~\ref{fig:correlfuncs}. The star-star auto-correlation functions
$C_{11}$ and $C_{22}$ (top
left panel) show a small signal on small scales that we attribute to some
over-correction of the DECam PSF in the areas near stars. The star-galaxy
correlation functions (bottom panels) show the same correlation on small
scales, with a magnitude several times smaller than the galaxy-galaxy
auto-correlation signal (which traces WL shear). The negligible value of the
$C_3$ in both star-star and star-galaxy pairs confirms that the PSF
circularization scheme introduces no major systematic error to shape measurement.

It should be noted that even after circularization, the measured ellipticity moments do not yet represent the true shapes of the galaxies. Both atmospheric seeing and the circularization of the image PSF make galaxies appear more round than they really are, and effectively dilutes the WL shear signal. We correct for this ``smearing" at the catalog level as detailed in \textsec\ref{sec:SourceSelection}. 

\subsubsection{Stacking\label{sec:Stacking}} 
\begin{figure*}
\begin{center}
\leavevmode
\includegraphics[width=4.5in, scale=1,clip=true]{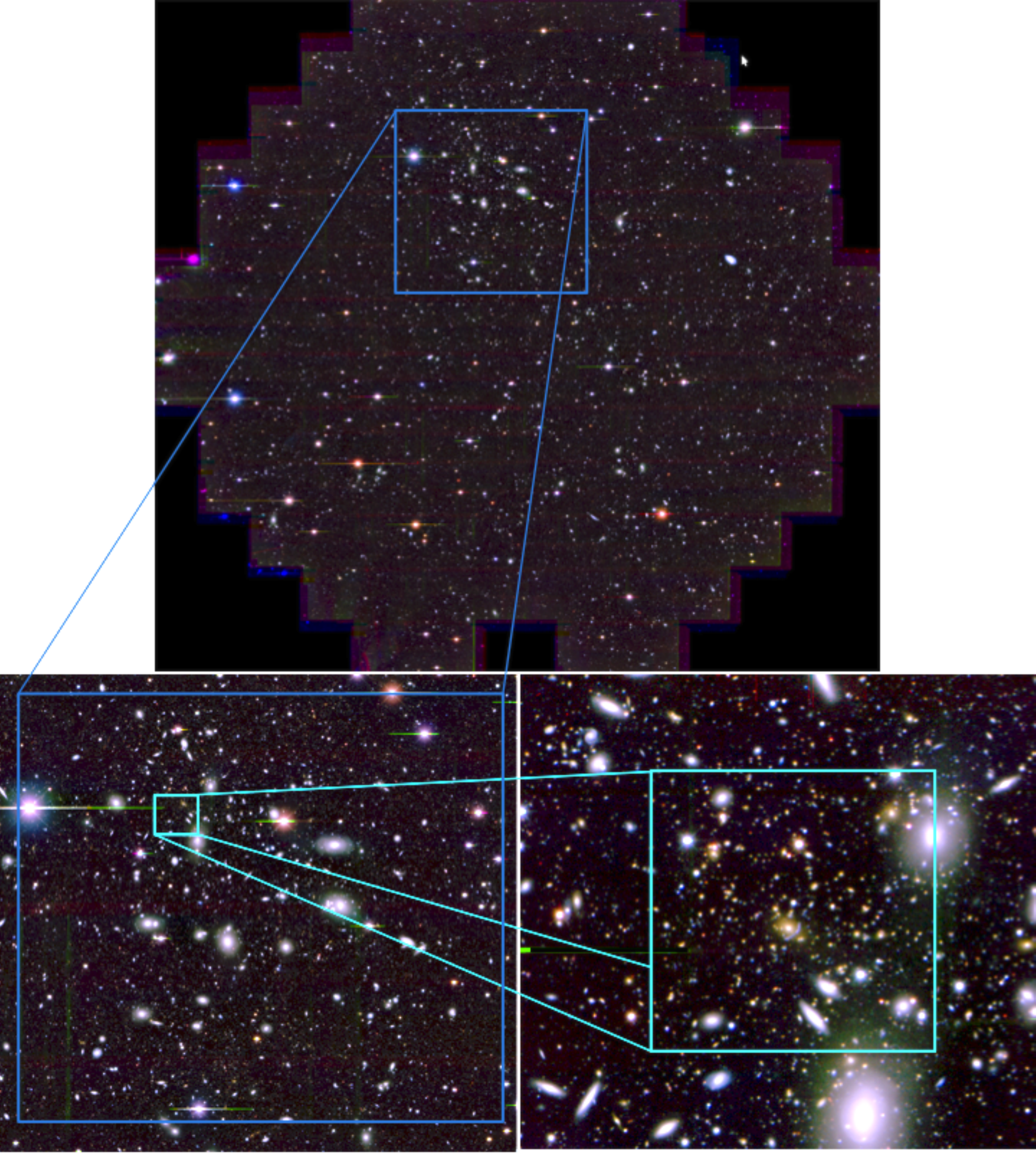}
\end{center}
\caption{\footnotesize{Sequence of $zrg$ composite images showing progressively higher magnifications of the Abell 3128 field. Top: The full DECam 1.5\degree x 1.5\degree field of view of the cluster and surrounding region. Left: Close-up view of the central 32' x 30' of Abell 3128. Right: 3' x 3' image showing the background cluster ACT-CL J0330-5227, which hosts SUMSS J033057-52281, a radio source at z = 0.44 (\cite{2007A&A...474..707W} and references therein). The image is centered on a strong gravitational lensing arc associated with this distant cluster.}}
\label{fig:3colorzoom}
\end{figure*}

Once we have corrected for CCD artifacts and, in the case of the $r$-band exposures, obtained polynomial fits to the PSF, we proceed to stack CCD exposures into a single image. Our procedure for stacking closely follows the one used in the the Deep Lens Survey; see \citealp{2006ApJ...643..128W} for full technical details. The steps undertaken to produce our stacked images are summarized here.
\begin{enumerate}
\item {\it Source Detection \& Characterization}. -- For each CCD image in the dataset, Source Extractor (\citealp{1996A&AS..117..393B}) is used to make a catalog of high $S/N$ objects. Source Extractor also generates the sky background-subtracted images that will be the final inputs to the final stack image. Subtracting the sky background at this stage eliminates the need to match sky levels at the stacking stage. At this stage, the ELLIPTO program (\citealp{2002AJ....123..583B}) is used to measure the so-called ``adaptive" second-order moments of objects in the CCD image. Adaptive moments are centrally weighted by an elliptical Gaussian, and their measurement is equivalent to finding the best-fit elliptical Gaussian for each object. Unlike SExtractor's intensity-weighted moments, which are computed within some limiting isophote, adaptive moments do not depend on magnitude. This property makes adaptive moments more advantageous for galaxy shape measurement and the identification of stars (\citealp{2006ApJ...643..128W}). 
%
\item {\it Star Identification}. -- As discussed in \textsection\ref{sec:PSF}, we use stars to trace out the spatial variation of the PSF across CCD chips. After Source Extractor catalogs have been made, stars are picked out for their membership in the respective stellar loci of size-magnitude and magnitude-surface brightness diagrams.  Initial identification proceeds automatically using an algorithm that identifies the typical stellar size, then selects the magnitude range for which there is a significant density enhancement at that size. To ensure an accurate sampling of the PSF for circularization, the $r$ band stellar catalogs are individually inspected and manually adjusted as needed. The stars used for $r$ band PSF fits are highlighted in the size-magnitude diagram of Figure~\ref{fig:sizemag}. There and elsewhere in the paper, size is defined as the sum of the second-order intensity moments $I_{xx} + I_{yy}$ with an additional factor $\rho^4$ to correct for non-Gaussianity~(\citealp{2002AJ....123..583B})

\item {\it Master Catalogs}. -- The DLS survey found that the MSCRED astrometric calibration of images is not good enough to stack them directly; small shifts between overlapping exposures would lead to spurious stretching of galaxy shapes~(\citealp{2006ApJ...643..128W}). To precisely define the astrometry of the final stack image, we match all the catalogs in equatorial coordinates to produce a master catalog. Every object that was observed in at least three exposures (within the tolerance of $1''.8$) has its mean right ascension, declination and magnitude recorded. Subsequently, the master catalog positions are used to define a coordinate system for the stack (a simple tangent plane projection with no optical distortion) and then transformed to pixel coordinates in the final stack image.

\item {\it Pixel Coordinate Transformations}. -- For every CCD in each exposure, matches between the Source Extractor catalog and the master catalog are used to define a transformation from CCD pixel coordinates to final stack image. 
For the $ugz$Y data not subject to WL analysis, less stringent astrometry requirements allow for the DLS default of a third-order polynomial. 
In the case of $r$ band exposures, the coordinate transformations 
must be defined by a fourth-order polynomial. Lower order polynomials underfit the variation at the edges of the CCD, resulting in a slight elongation of galaxies in the final stack image and ultimately leading to bands of spurious shear signal in the WL convergence maps. However, many CCDs from the edge of the exposure ($\sim 20\%$) have too few matches with the master catalog to support a 4th order fit and must be excluded from the final stack image.  

\item {\it PSF Circularization}. -- Once the pixel transformation polynomials are determined, we generate PSF circularization profiles for every CCD image in the dataset. By default, the adaptive moment combinations $I_{xx} - I_{yy}$, $I_{xx} + I_{yy}$ and $I_{xy}$ are fit automatically with 4th order polynomials. The exceptions are PSF profiles for the $r$-band imaging, which are prepared in advance using the multi-stellar catalog procedure laid out in \ref{sec:PSF}. 
\begin{figure}[h]
	\includegraphics[width=0.450\textwidth]{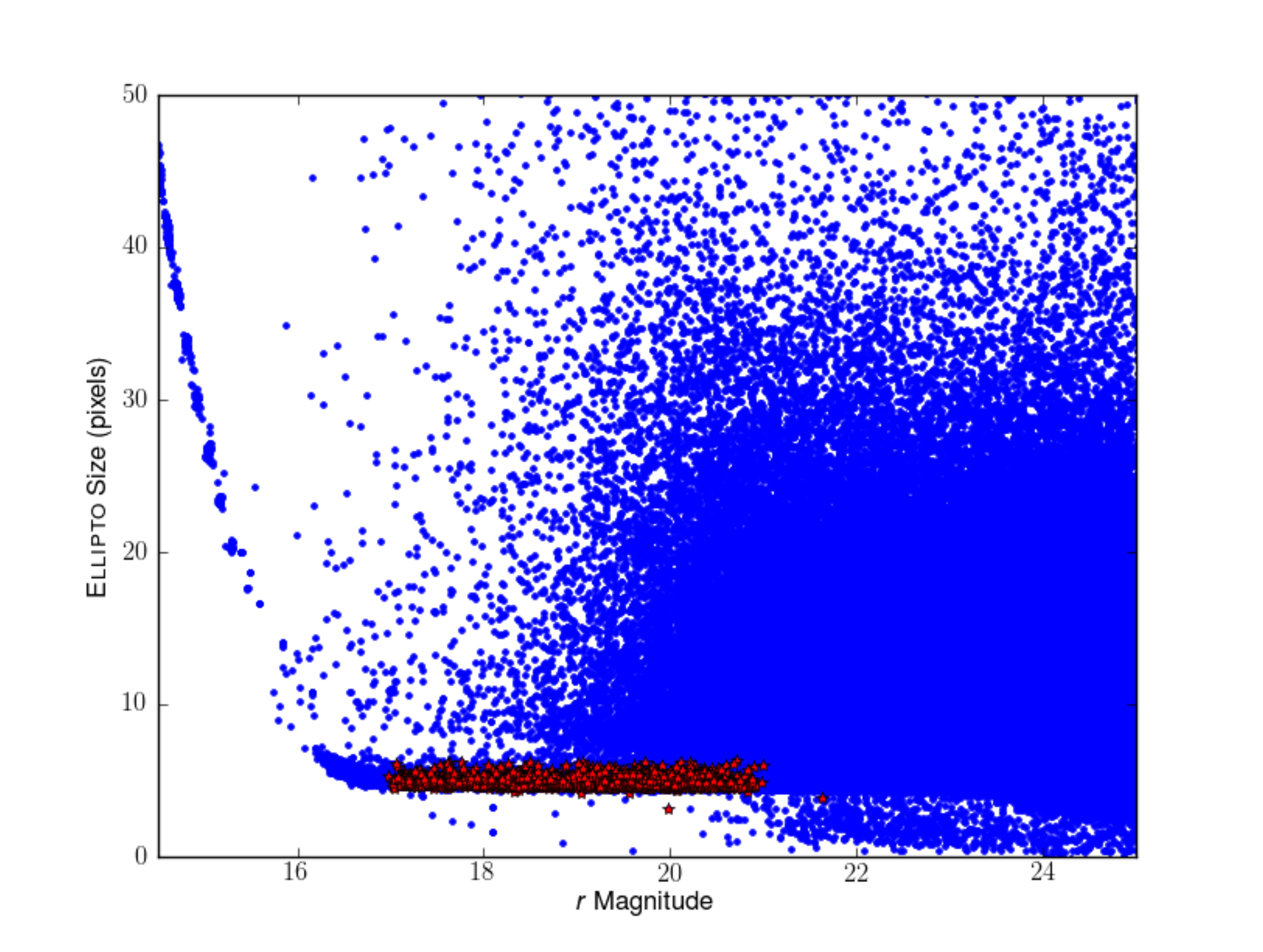}
	\caption{\footnotesize{Size-magnitude diagram of the object catalog generated from the final stacked $r$ image. At this stage, the catalog is filtered for objects with Source Extractor or ELLIPTO error flags, but no other cuts are applied. The stellar locus is the stripe of objects with size $\sim 5$ and $15.5 < m_r < 21$; red stars mark objects used in our multi-catalog PSF circularization procedure. Objects in the ``second stellar locus'' with sizes around 2 pixels and $m_r<23$ are pixel noise variations. Spurious detectections along the bleed trails of saturated stars cause the uptick at $m\sim16$.}}
	\label{fig:sizemag}
\end{figure}

\item {\it Photometric Calibration}. -- Before CCD exposures are stacked, their photometry is calibrated to ensure consistent object magnitudes everywhere in the final stack image.  This is made more difficult since, due to the curvature of the focal plane, pixels at the edges of the DECam CCD subtend more sky area than pixels at the center of the array.  
To correct for focal plane distortion, catalog magnitudes are multiplied by the Jacobian of the pixel coordinate transformations computed in step 4 and then gathered into a master photometric catalog. For each CCD exposure, we then derive a relative photometric offset by matching its catalog to the master photometric catalog and computing the $3~\sigma$ clipped mean of the magnitude differences of the matching objects.
\end{enumerate}
Finally, to produce the stacked image, we implement the Deep Lens Survey's DLSCOMBINE algorithm. For each pixel in the output image, DLSCOMBINE loops over contributing pixels in the input images and applies the relevant bad pixel masks, PSF circularization kernels, coordinate transformations and photometric offsets. A $3~\sigma$ clipping is applied before the mean pixel value is returned. A 3-color composite image made from the $z$, $r$ and $g$ stacked images is shown in Figure~\ref{fig:3colorzoom}. 

\subsection{Source Selection\label{sec:SourceSelection}}

We produce a source catalog from the final stacked $r$ image using Source Extractor~(\citealp{1996A&AS..117..393B}). The Source Extractor detection significance and deblending thresholds are deliberately set to low values so that the faint background galaxies on which we perform WL analysis will be recognized. To clean out the accompanying multitude of junk Source Extractor detections, we perform a number of cuts to the initial object catalog. All detections with high Source Extractor and ELLIPTO error flag values may easily be filtered out, 
but ridding the catalog of pixel noise ``detections" presents a special problem. Because their few counts are contained within a small isophotal area, SExtractor frequently assigns them reasonable magnitudes, and measurement of their adaptive moments produces no error flags.  To filter out such pixel noise from the object catalog, we removed detections with ELLIPTO-determined fluxes of less than 100 counts. The cuts on error flags decreased the initial object catalog of 1.4 million objects by about 22\%, and the flux cut decreased it by a further 43\%. 

The object catalog is then subject to the the size and magnitude cuts typical of weak lensing studies, which must remove stars, low-redshift galaxies and noisy sources while maintaining a large sample. The criteria for inclusion in the final sample were isophotal {\em r} magnitudes between 17.2 and 24.6, and object size between 6.0 and 200 pixel$^2$, where size was defined in~\ref{sec:PSF}. The atypically generous upper limit of the size cut reflects the fact that A3128 is at very low redshift, and its ``background" contains many large galaxies. Requiring that objects be detected in all four bandpasses {\em de facto} constitutes an additional catalog cut, eliminating 30,000 objects from the final catalog. At this point, the filtering has reduced the catalog down to 200,000 sources in total (25 sources arcmin$^{-2}$).  

\begin{figure}[tb]
 	\centering
	\includegraphics[width=0.5\textwidth,clip=true]{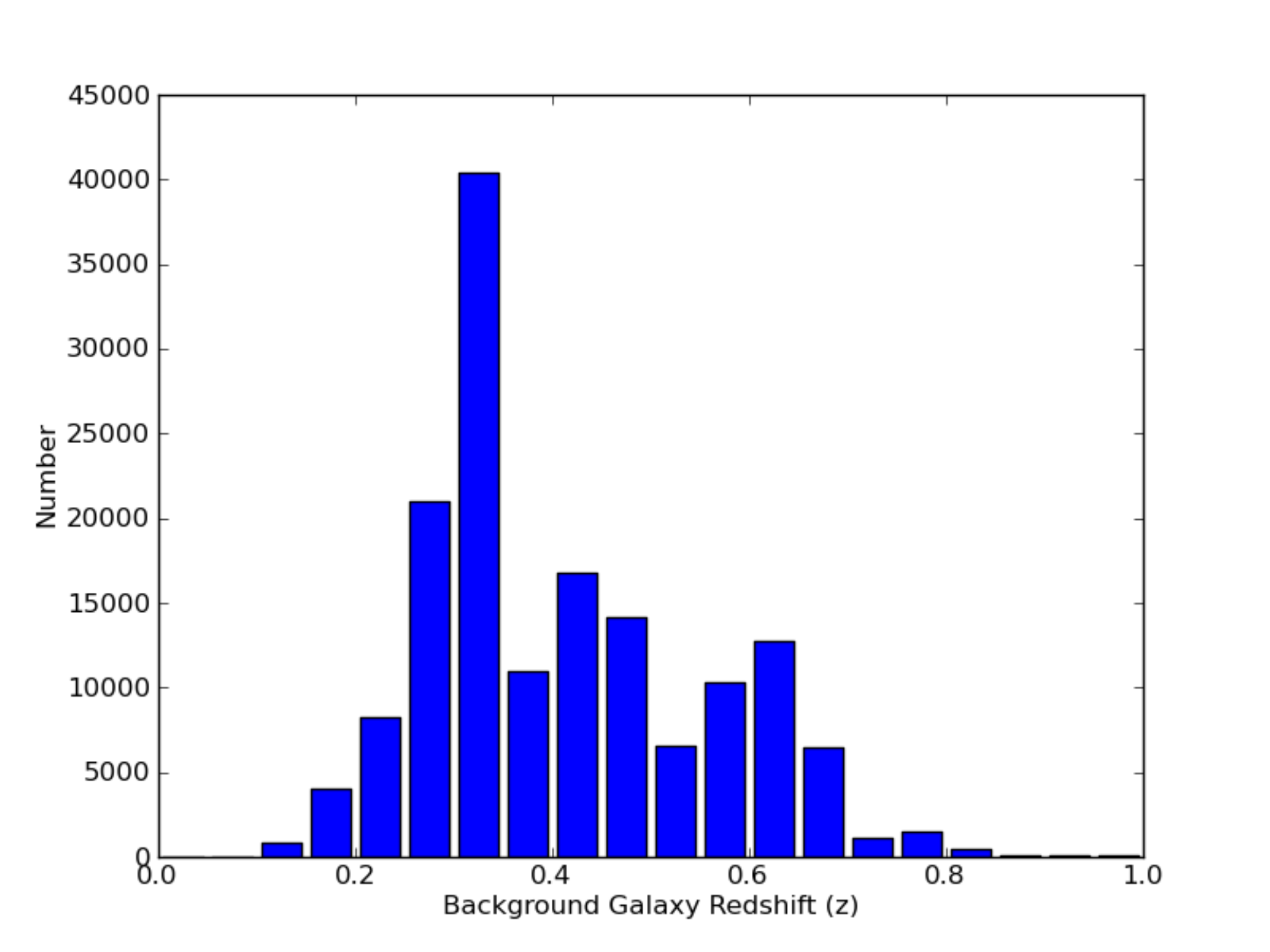}
	\caption{\footnotesize{Redshift distribution of the final catalog, filtered for low-redshift galaxies. The majority of our sources are at $z\sim0.30$, with secondary peaks at $z\sim0.46$ and $z\sim0.60$.}}
	\label{fig:redshift}
\end{figure}
Any galaxies in the foreground of A3128 will not be sheared by the cluster, and so their presence in the final source catalog dilutes the convergence measured on the lens. To filter out low redshift contaminants, we derived photometric redshifts using BPZ~(\citealp{2000ApJ...536..571B}) with the standard HDF prior. The $ugrz$ magnitudes of cataloged objects were submitted to the program, although since the u-band observations are rather shallower than the $grz$ observations, the photometric redshifts are essentially three-point fits. We used the default CWWSB template set (E, Sbc,
Scd, Irr, SB3, and SB2) with no modifications, and allowed three levels of interpolation between neighboring templates. The range to be considered was restricted to $0.03 < z_{BPZ} < 3.0$. To evaluate BPZ results on low-redshift galaxies, we identified 10 galaxies in the observations with spectroscopic redshifts $z \sim 0.06$ and used BPZ to determine their photometric redshifts. This test calls attention to the uncertainty in BPZ results, as all 10 galaxies were assigned redshifts between 0.07 and 0.13. This range is comparable to the per-galaxy rms error found by other studies (e.g. \citealp{2008ApJ...673..163S}). While most galaxies that are truly in the foreground of A3128 are large enough to be eliminated by the size cuts described above, we nonetheless took the low-redshift uncertainty into account when developing our redshift selection criterion of z$_{BPZ} > 0.19$ at greater than 95\% probability. About 169,000 objects (21 sources arcmin$^{-2}$) remain after this latest cut is made.

One last step remains before we can make convergence maps for Abell 3128. Even after circularization, atmospheric seeing and the circularization procedure itself still smear out the measured adaptive moments, making galaxies appear rounder than they really are and diluting the WL shear signal. As a correction, each galaxy's ellipticity is divided by a factor $R$ which relates the size of the galaxy to the mean stellar PSF size (\citealp{2002AJ....123..583B}). Galaxies must have a measured $R > 0.2$ for the correction to be successful, otherwise they are cut from the catalog. Only about 5\% of galaxies remaining in the catalog at this stage fail to meet that criterion, which leaves a final source density of 20 sources arcmin$^{-2}$.
\begin{figure}[ht]
\begin{center}
	\includegraphics[width=0.48\textwidth,clip=true]{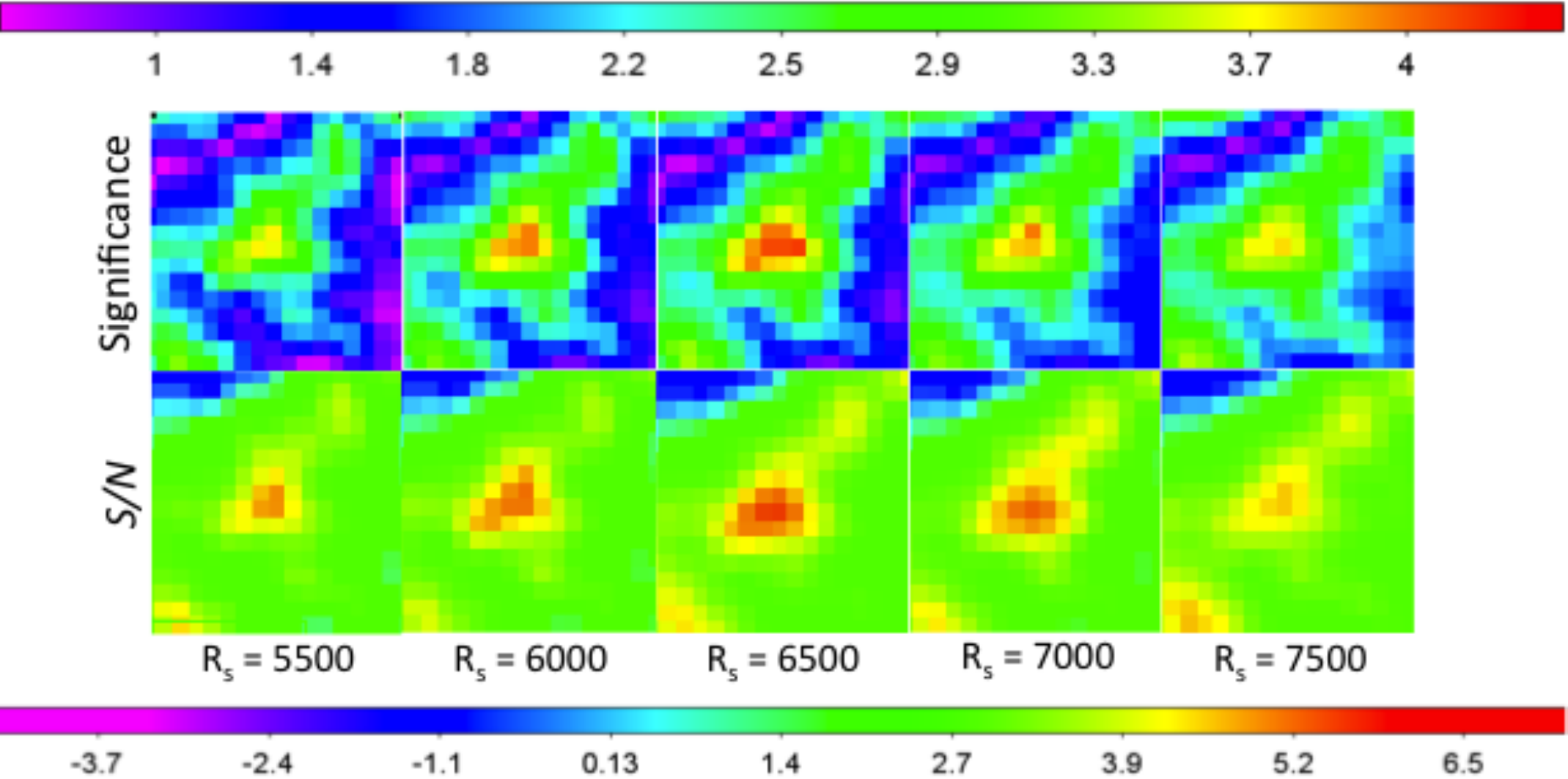}	
\caption{\footnotesize{An aperture mass peak in five maps of consecutively larger filter radii. {\em Top:} Detection significance of aperture mass in units of $\sigma$. {\em Bottom:} Signal-to-noise of the aperture mass. In this example adapted from our A3128 observation, the signal peaks in both significance and signal-to-noise at a Schirmer filter radius of 6500 pixels.}}
\label{fig:sigmashear}
\end{center}
\end{figure}
\subsection{Convergence Mapping and Quantified Detection of Substructure\label{sec:Convergence}}

The catalog finally contains the true shapes of confirmed galaxies behind A3128, and may be used to reconstruct the cluster's projected mass. 
To extract the aperture mass signal of A3128 from the tangential ellipticities of background sources, this study relies on software developed by co-author Huwe as part of his thesis work (\citealp{huwethesis}, Huwe \& DellÕAntonio 2014, in preparation). Our particular implementation of this software is presented here. 

Our software first bins the A3128 image into blocks 200 pixels on a side to reduce computation time. Using the adaptive moments of galaxies in the catalog, the software then returns the filtered aperture mass statistic (Equations \ref{aperturemassstat} and \ref{Schirmerfilter}) within each 200-pixel block. To quantify our mass reconstruction of the A3128 observation, we construct a WL signal-to-noise map as follows. Random noise maps are generated by recalculating the aperture mass statistic (with the same Schirmer filter and pixel block sizes) on a catalog of shuffled galaxy positions and moments. The random maps will initially be dominated by non-Gaussian shape measurement error, so the randomization process is repeated 100 times. Assuming that the errors in the aperture mass reconstruction will then be Gaussian, the variance of each image block in the random maps represents the $1~\sigma$ noise level of the $M_{ap}$ statistic. Dividing the $M_{ap}$ signal map by the variance of the random maps, pixel block by pixel block, yields an estimate for the signal-to-noise. 

Because WL distortion is tangential to the direction to the center of mass, an image should have no systematic B-mode (curl-like) distortion in the shapes of background galaxy. The lensing signal should thus vanish when $e_{\tan}$, the E-mode (curl-free) component of shear, is replaced with the B-mode component $e_c$ defined in Equation \ref{ec}. Any significant WL peaks obtained when $e_{\tan}$ is replaced with $e_c$ in Equation \ref{aperturemassstat} would not come from the cluster, but instead indicate some systematic error in the analysis. Since most systematics are expected to add equal power to E- and B-modes~(\citealp{2003AJ....125.1014J}), we generate B-mode signal-to-noise maps to control for bias in our analysis. 

Both E- and B-mode signal-to-noise maps treat the errors in the convergence field as Gaussian, but this assumption may not be warranted. To further quantify confidence in the results of our mass reconstruction, we calculate the detection significance of features in the signal-to-noise map with the following algorithm.  Using the same filtered aperture mass statistic as before (Equation \ref{aperturemassstat}), the software creates a signal file and then iterates through some large number of random noise reconstructions which are stored in memory. At every 200-pixel block of the observation, the software tallies how many noise reconstructions had a greater magnitude of WL signal than the signal file. This number should be close to 0 for blocks near the cluster center, but in massless regions of the observation will be roughly 50\% of the total number of random iterations. When inserted into an inverse cumulative distribution function, this number is converted into a Gaussian-type confidence $\sigma$ which quantifies the significance of the shear signal in that pixel block. The maximum attainable $\sigma$ will depend on the number of noise iterations; our software generates 100000 random maps which corresponds to a maximum confidence of $4.42~\sigma$.

  In addition to a magnitude given by its $\sigma$ value, the software assigns to each significance map pixel the sign of the corresponding pixel block in the original signal map. Hence, regions in the significance maps with negative values correspond to statistically significant underdensities in the mass distribution, while positive $\sigma$ means an area of mass enhancement compared to the mean. 

To search for mass concentrations, the significance maps are thresholded above $+3.8 \sigma$, and potential substructure peaks are identified by inspection. For each group of contiguous image blocks with significance greater than $+3.8~\sigma$, we follow the feature through a range of Schirmer filter scales. 
Relatively small Schirmer filter radii $x_c$ do not encompass all the shear signal from the feature, whose significance will subsequently be suppressed. As the filter radius increases, the significance of the detection increases before peaking at some Schirmer filter size which is then the characteristic scale of that substructure. Further filter expansions eventually lead to the merging of the substructure signal into the overall cluster signal. An example of this increase and decrease in significance is shown in Figure~\ref{fig:sigmashear}. 

\subsection{NFW Shear Profile Fitting\label{sec:NFWfitting}}

Recalling that aperture mass maps return only the relative mass enhancements
in an observation, we have written software which fits A3128 and its substructures
with axisymmetric NFW weak lensing shear profiles to constrain their physical
masses.  Different algorithms are employed to fit the observation with single and multiple NFW halos, but in both cases the software first obtains the scaling factor
$\Sigma_{cr}$ (Equation~\ref{sigmacrit}) for each galaxy in the catalog. Requisite angular diameter distances for $\Sigma_{cr}$ are computed from the galaxies' BPZ
redshifts. The software then varies the $M_{200}$ of an NFW halo and computes
the corresponding $r_{200}$ under a {\em Planck} XVI cosmology. The halo's concentration $c$ is obtained by inserting its $M_{200}$ into the
empirical relation of  \cite{2013ApJ...766...32B} for their full cluster
  sample. The $r_{200}$ and $c$ parameters become the $r_s$ and $\delta_c$ which characterize
an NFW halo's mass distribution and hence its shear profile. 

To fit a single NFW mass to the primary WL peak of A3128, as  $M_{200}$ is varied the software follows the prescription of \cite{2000ApJ...534...34W}
to compute the halo's reduced shear at the location of every background galaxy. We find the best-fit $M_{200}$ by using the
parabolic extrapolation method of \cite{2007NR} to minimize $\chi^2$ residuals between the
NFW halo's shear profile and the $e_{\tan}$ measured on the image.  

Simultaneous fitting of NFW masses to multiple substructures requires a different approach: the (tensor) reduced shear from multiple NFW
peaks does not add linearly, so the prescription of \cite{2000ApJ...534...34W}
is not directly applicable. Instead, we use the fact that background galaxies
experience an individual displacement $\vec{\beta}$ from each NFW halo. 
In the weak lensing approximation, the displacements from multiple NFW haloes add linearly: $\vec{\beta}_{tot} =
\sum{\vec{\beta_i}}$. The total shear and convergence at every point in the
image can then be built from derivatives of the Jacobian ($\partial
\vec{\beta}_{tot}/\partial \vec{\theta}$) using the formulae of
\cite{2002A&A...390..821G}. Presupposing the locations of their centers have been established, we vary
the $M_{200}$ of NFW haloes centered on each substructure, obtain the
corresponding reduced shear at the location of every background galaxy, and minimize the profiles' $\chi^2$ residuals
against the galaxies' tangential ellipticities. 

We emphasize that neither the single-peak nor multiple-peak mass estimates in this work result from fitting 1-D NFW shear profiles to azimuthally averaged galaxy ellipticities, though such an approach is common in the literature. Rather, our NFW fitting method uses the full positional information of every galaxy in the catalog. Tangential shear profiles shown below (Figure~\ref{fig:shearprofile}) are for illustrative purposes only. We also note that in all mass estimates, NFW shear profiles are centered on the aperture mass peak's highest
signal-to-noise pixel in convergence maps. However, due to our binning scheme
(cf.~\textsec\ref{sec:Convergence}), each convergence map pixel actually spans 200
pixels ($53''$) on the observation. The resultant ambiguity in the
identified center of a WL peak could bias mass estimates through a mis-computation of the galaxy ellipticity; we investigate this potential centroid bias in \textsec\ref{sec:nfwfits}.


\section{RESULTS\label{sec:Results}}
\subsection{Detection of Primary Cluster Aperture Mass\label{sec:OverallDetection}}

\begin{figure*}[tbp]
 	\centering
        \includegraphics[width=0.80\textwidth,clip=true]{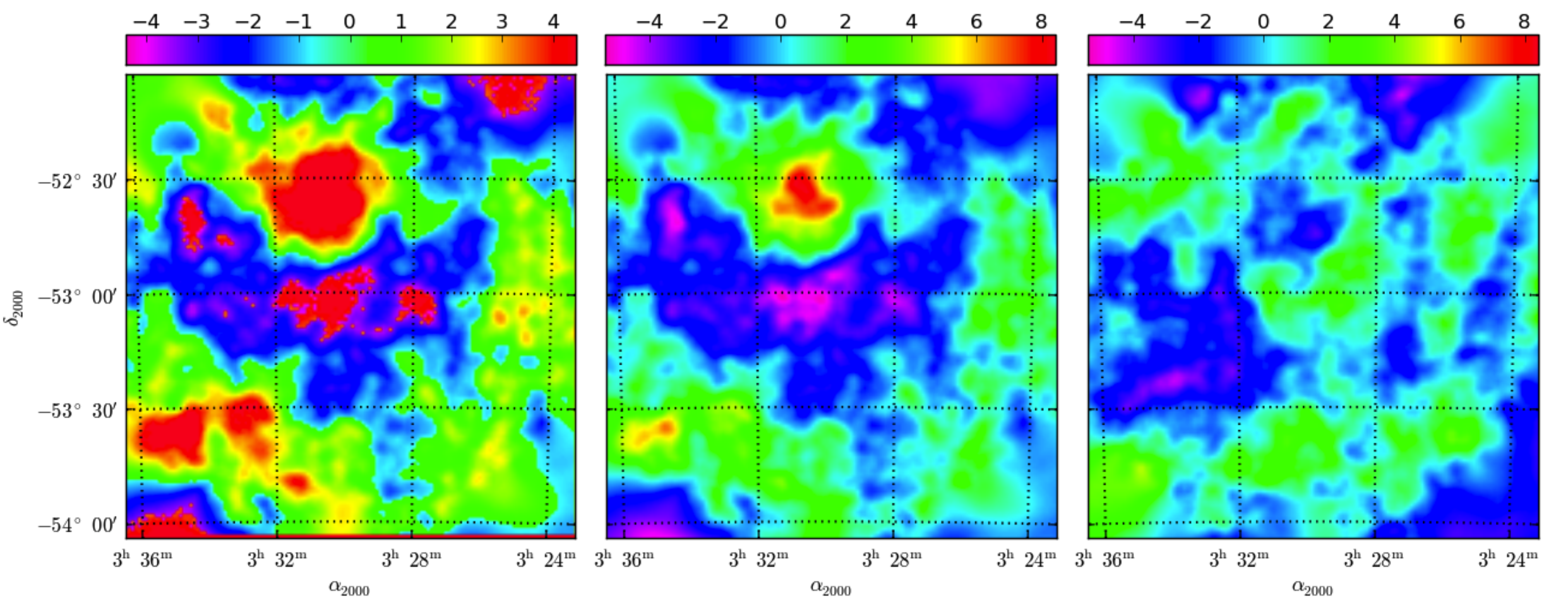}
	\caption{\footnotesize{Reconstructed weak lensing convergence maps for Abell
            3128, all with 52''/pixel resolution on the observation. {\em Left:} Significance map 
            made using a Schirmer aperture of radius 10000 pixels (= 44'). The
            primary peak is at the top left, and saturates our significance at
            $4.42 \sigma$. {\em Center:} Signal-to-noise map  and 44' Schirmer aperture, which yielded the maximal cluster $S/N$ of 8.4 at two separate locations. {\em Right:} B mode  {\em S/N} map for 10000 pixel aperture radius (primary peak filter size). }}
	\label{fig:A3128convergence}
\end{figure*}

Applying the procedure laid out in \textsec\ref{sec:Convergence} to the observation, we report the detection of the Abell 3128 weak lensing signal at high significance. The top left of Figure~\ref{fig:A3128convergence}(a) shows the primary aperture mass peak, which saturates our significance maps with $\sigma > 4.42$ at all Schirmer filters larger than 6000 pixels. To identify the aperture size which best characterizes the cluster, we constructed signal-to-noise maps using Schirmer filters up through $R_S = 14000$ pixels. Within this range, the A3128 aperture mass achieves its peak $S/N$ of 8.4 at two distinct locations with an aperture size of $R_S = 10000$ pixels. Since the Schirmer filter weight peaks sharply at $\sim0.10R_S$, the primary A3128 signal spans roughly 4.4' on the image. 

From the location of the highest signal pixels in the $S/N$ map of Figure~\ref{fig:A3128convergence}(b), we might assign the primary weak lensing peak of Abell 3128 to coordinates $\alpha = 3^h 30^m 50^s.5, \delta = -52\degree31'15''$. However, the presence of a massive high-redshift cluster (visible in Figure~\ref{fig:3colorzoom}) only 6' from the A3128 X-ray center confounds the location of its center of mass. Instead, we defer this question to \textsection\ref{sec:hiz}, where the WL signal of the background cluster is probed.

 In addition to the primary cluster peak, several other high-$\sigma$ aperture masses appear in Figure~\ref{fig:A3128convergence}: the $S/N \sim 7$ clumps at the bottom left of Figure~\ref{fig:A3128convergence}(b) also saturate significance maps with $\sigma=4.42$. These clumps are better characterized at small Schirmer filter sizes, as described in in \textsec\ref{sec:substructure}. Immediately below A3128, the WL maps show a highly significant void in the local dark matter distribution.

{
}
\begin{figure*}[tbp]
 	\centering
	\includegraphics[width=0.65\textwidth,clip=true]{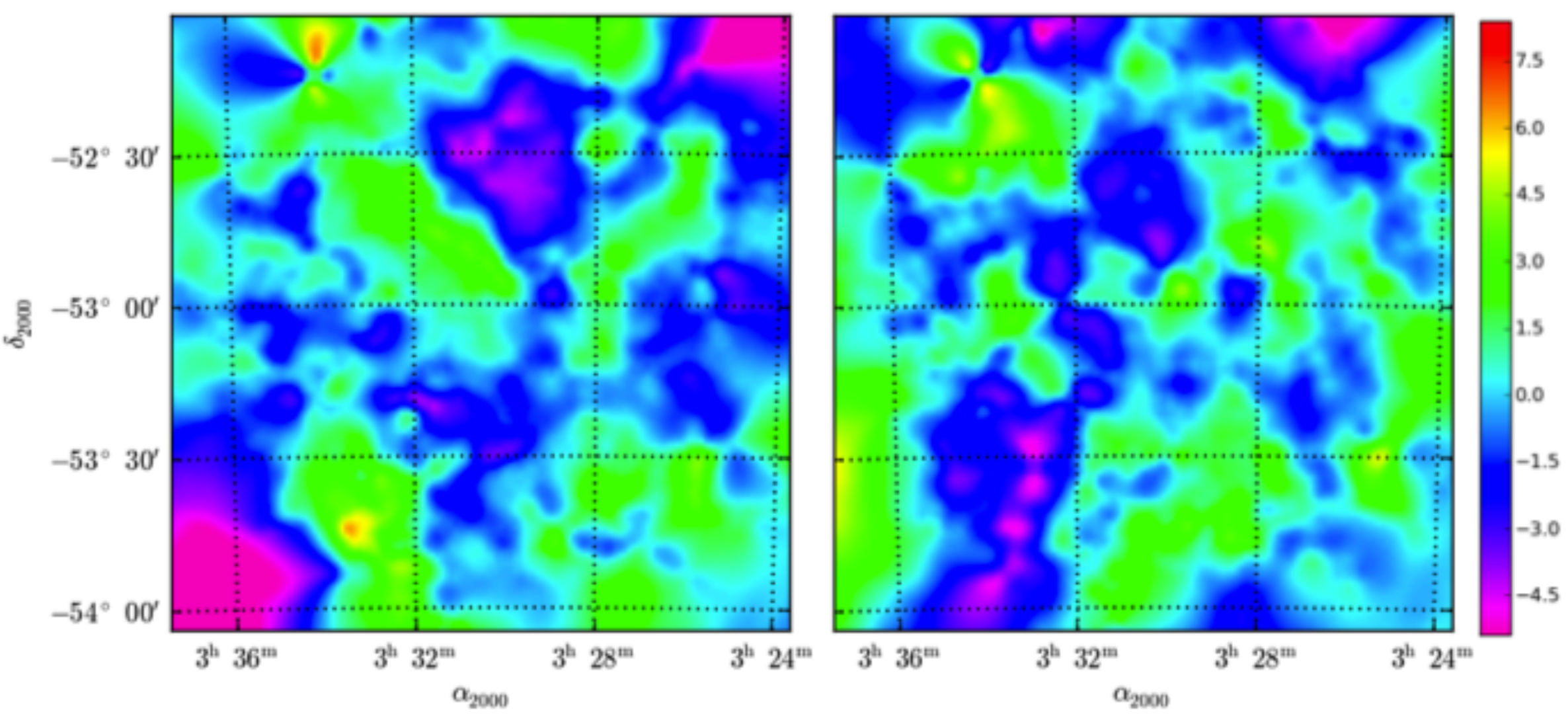}
	\caption{\footnotesize{Reconstructed aperture mass maps with 52'' resolution made from the catalog
            of stars used to circularize the DECam PSF . {\em Left:}
            Signal-to-noise map made with a Schirmer aperture of radius 10000 pixels (= 44'), the filter size of the A3128 peak.  {\em Right:} B mode  {\em S/N} map for 10000 pixel aperture size. Note that both maps have significant areas of negative WL signal, indicating a net tendency of background galaxies to be alined radially outwards.)}}
	\label{fig:PSFstarSN}
\end{figure*}

A signal-to-noise map made with the B-mode statistic $e_c$ (Equation~\ref{ec}) controls for any systematic error in the galaxy catalog (\textsection\ref{sec:Convergence}). Accordingly, the B-mode signal-to-noise map for the $R_S = 10000$ pixel aperture is shown in Figure~\ref{fig:A3128convergence}(c). This map shows a marked anti-correlation between B-mode and E-mode signal in the neighborhood of the cluster, which becomes even more noticeable at small aperture sizes like Figure~\ref{fig:A3128substructure}. We investigate by making E- and B-mode maps from the catalog of stars used to circularize the PSF. In the absence of a systematic error, WL reconstructions from a catalog of perfectly circular(ized) stars would yield signal-to-noise maps indistinguishable from random noise. Instead, both the E- and B-mode $S/N$ maps of Figure~\ref{fig:PSFstarSN} display a negative signal near the cluster center. Given that systematic errors should add equal power to E- and B-modes, it is expected to see the negative signal in both the B-mode $S/N$ map of Figure~\ref{fig:A3128convergence}(c) and the E- and B-mode $S/N$ maps of Figure \ref{fig:PSFstarSN}.
\LongTables
\begin{deluxetable}{cccc}
	\tabletypesize{\scriptsize}
	\tablecaption{\footnotesize{Centroids from Previous Studies}\label{CentroidCoords}}
	\tablewidth{0pt}
	\tablehead{
	\colhead{} & \colhead{$\alpha$} & \colhead{$\delta$} &\colhead{}\\
	\colhead{ID} & \colhead{(J2000.0)} & \colhead{(J2000.0)}&\colhead{Reference}
	}
	
		\startdata
 			A3128 Optical Center & $3^h 30^m 43^s$ & $-52\degree31' 50''$&Rose et al. (2002)\\
			A3128 X-ray Center  & $3^h 29^m 40^s$ & $-52\degree28' 50''$ &Werner et al. (2007)\\
			X-ray ``SW Peak" & $3^h 29^m 55^s$ & $-52\degree34' 50''$&Werner et al. (2007)\\
			ACT-CL J0330-5227 & $ 3^h 30^m 57^s$ & $-52\degree28' 14''$&Menanteau et al. (2010)\\
		\enddata

\end{deluxetable}%

Equation~\ref{ec} shows that negative WL signal is caused by a net tendency of objects in a region to be aligned radially outwards. Furthermore, the DECam PSF is known to have a strong radial component (cf. the ``pincushion" of Figure~\ref{fig:shearsticks}). Hence, this negative signal is attributable to an undercorrection of the PSF ``pincushion" near the bright galaxies of the cluster, likely due to gaps in stellar coverage. This interpretation is supported by the weakening of the effect at large Schirmer filter radii, where the inclusion of more galaxies at random orientations washes out any localized residual correlations in the PSF. 

We constrain the magnitude of this systematic effect as follows. We amalgamate the star and galaxy convergence maps and tally up WL signal enclosed in a circle centered on the cluster. This value is then compared to the equivalent signal in the convergence map made from galaxies alone.  We find that the combined stars/galaxies convergence map has 5\% more power than the galaxies-only signal map, likely because the systematic undercorrection of the PSF subtracted away some of the cluster's original convergence signal. Since WL shear and mass scale linearly, we expect that the NFW fits presented in Table~\ref{NFWfits1} underestimate the true mass by $\sim 5\%$.

\subsection{High-Redshift Background Cluster\label{sec:hiz}}


\begin{figure*}[htb]
 	\centering
	  \includegraphics[width=0.95\textwidth,clip=true]{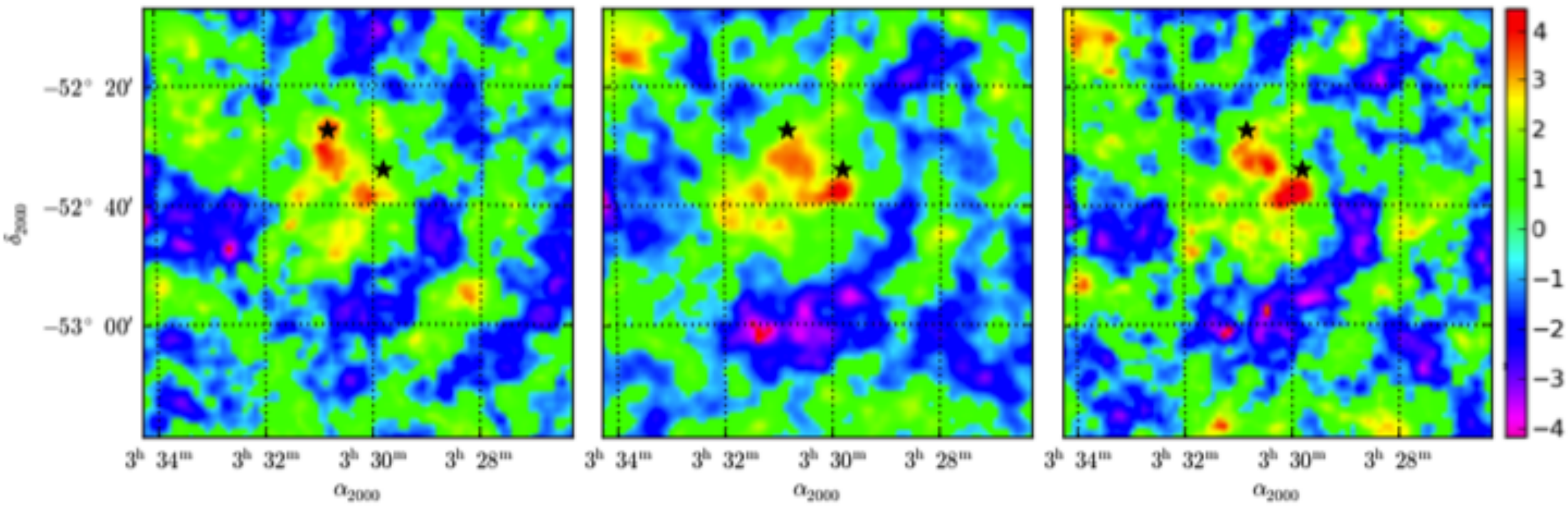}
	\caption{\footnotesize{Significance maps marked with the published positions of the Abell 3128 X-ray ``southwest peak" (bottom right star) and the $z=0.44$ cluster ACT-CL J0330-5227 (top left star).  Pixels in all maps span 52'' on the observation. The two extended high-$\sigma$ regions apparent in all three panels are the principal substructures of A3128; they are discussed as A1 and A2 in \textsec\ref{sec:substructure}. {\em Left:} Close-up of map made with background galaxy redshift restricted to $z \ge 0.44$ and a Schirmer aperture of 4000 pixels. The high-redshift cluster is plainly visible. {\em Center:} Map made with galaxies at redshifts between $0.16 \le z \le 0.4$ and a 6000 pixel aperture. While A3128 is still identifiable, albeit at slightly reduced significance, the high-redshift cluster has dropped out of view. {\em Right:} Map made with the full background galaxy sample and a 4000 pixel aperture.}}
	\label{fig:hiz2}
\end{figure*}

Visible in the last panel of Figure~\ref{fig:3colorzoom} is an interloping high-redshift cluster of galaxies, complete with the blue arc that is the hallmark of strong gravitational lensing. Multiple studies have confirmed that this background ``cluster behind a cluster," identified in the literature as ACT-CL J0330-5227, lies at $z=0.44$ and is the source of the northeastern lobe of X-ray emission in A3128~(\citealp{2007A&A...474..707W}). All cataloged galaxies with $z > 0.44$ will bear lensing signal from both clusters, and so the higher-redshift cluster must be precisely characterized to prevent confusion with substructure in A3128.

To disambiguate the signals of A3128 and ACT-CL J0330-5227, we prepared two subsamples using BPZ redshifts and probabilities:  an intermediate-redshift sample containing galaxies at $0.19 < z < 0.4$ with greater than 75\% probability, and a high-redshift sample containing galaxies at $z >0.443$ with greater than 75\% probability. The intermediate-redshift catalog contains 75,774 galaxies at a mean redshift of $z = 0.31$, and the high-redshift catalog contains 52,847 galaxies at a mean redshift of $z=0.68$. The intermediate-redshift galaxies experience WL distortion only from A3128, since they lie behind A3128 but in front of ACT-CL J0330-5227; the high-redshift galaxies are behind both clusters and experience distortion from both. 

In Figure~\ref{fig:hiz2}, significance maps made with the two redshift subsamples are compared to the the full galaxy sample map. The X-ray position of cluster ACT-CL J0330-5227 and the ``southwest peak" of A3128 X-ray emission are marked with black stars. Using the high-redshift galaxy subsample (left panel), we detect at high confidence the WL signal from ACT-CL J0330-5227 and both the A1 and A2 substructures of A3128. In the reconstruction with the intermediate-redshift galaxy subsample (center panel), the WL signal of A3128 is still distinct but the high-redshift cluster has dropped out of view. A reconstruction made with the full galaxy sample is shown in the right panel of Figure~\ref{fig:hiz2}; the inclusion of a large number of galaxies at $z < 0.44$ dilutes the signal of ACT-CL J0330-5227 and makes it harder to discern. The fact that the background cluster does appear or disappear with such selections bolsters our confidence in the galaxies' photometric redshifts, and by extension the angular diameter distances required to fit NFW profiles to aperture masses. 


Based on the absence of high-$z$ cluster signal in Figure \ref{fig:hiz2}(b), the intermediate-redshift galaxy subsample may be used to authentically identify the A3128 barycenter. From the location of the highest $\sigma$ pixel in large-aperture significance maps, we report the primary weak lensing peak (and presumably barycenter) of the cluster at  $\alpha = 3^h 30^m 16^s.5,\delta = -52\degree33' 57''$. This WL center is offset by 4.7 arcminutes from the optical center of the galaxy distribution, but by 6.24 arcminutes from the published X-ray center. Instead, the WL potential center coincides more closely with the ``southwest peak" of X-ray emission. Having established a location for ACT-CL J0330-5227 and the A3128 barycenter in our WL reconstructions, we may proceed in identification of substructures in A3128 with more assurance.

\subsection{High Significance Substructure\label{sec:substructure}}

\begin{figure*}[htb]
 	\centering
        	\includegraphics[width=0.65\textwidth,clip=true]{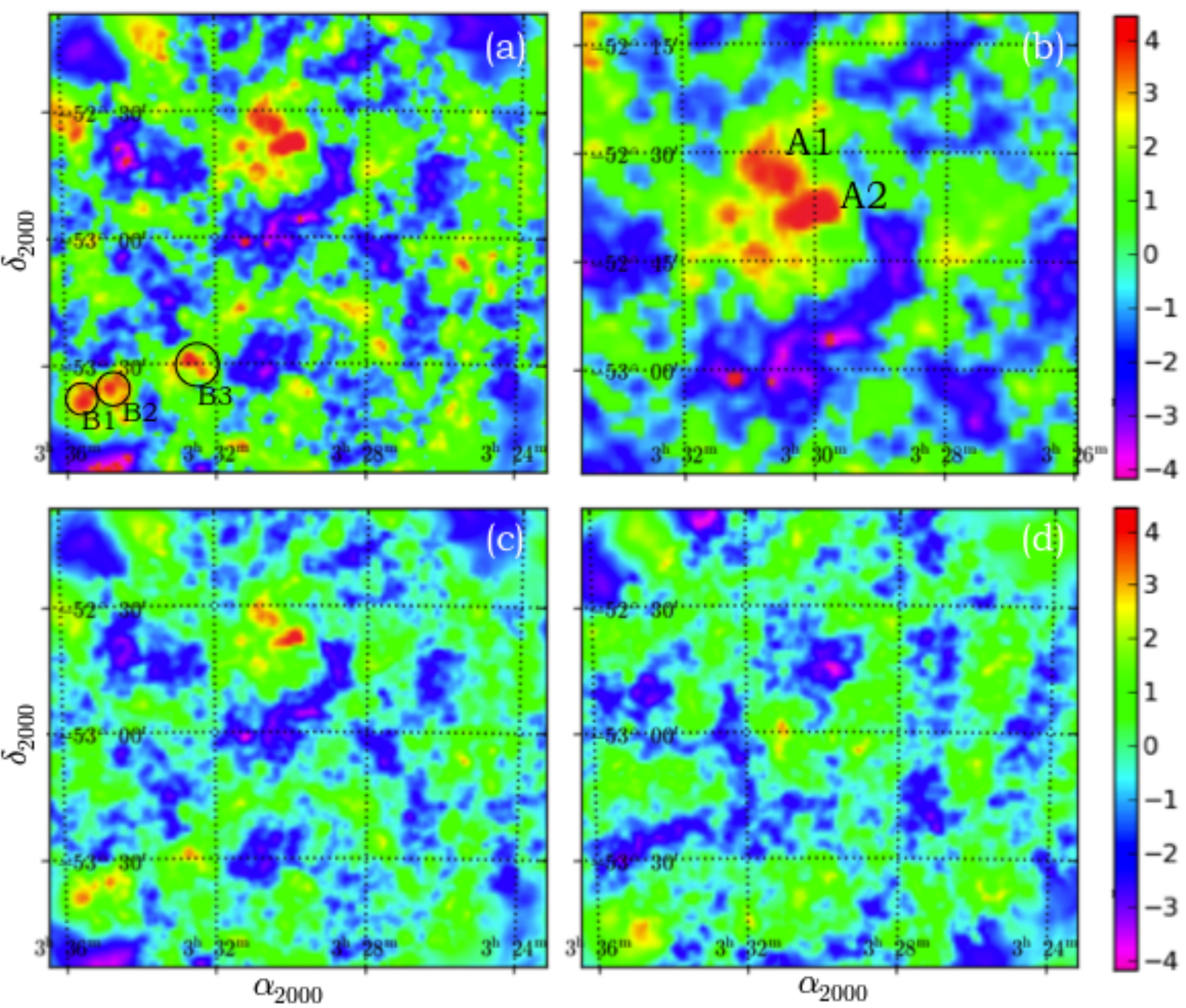}
	\caption{\footnotesize{High resolution convergence maps made with with 52''/pixel resolution on the A3128 observation. {\em Top Left:} Significance map made with a Schirmer filter of $R_S = 4500$ pixels ($= 19'$), in which the two substructures attain 4.42 $\sigma$ detection significance.  Several peaks to the bottom left (B1, B2 and B3) also attain high significance. {\em Top Right:} Magnified view of significance map at left, highlighting central cluster. {\em Bottom left:}  E-mode signal-to-noise map made with a Schirmer filter of $R_S=4500$ pixels. The A1 and A2 substructure peaks are detected at $S/N>5.6$. {\em Bottom Right:} B mode  {\em $S/N$} map made with $R_S = 4500$ pixel aperture.}}
	\label{fig:A3128substructure}
\end{figure*}

Our significance maximization procedure yields two substructures within A3128 proper, which are visible in the left-hand panel of Figure~\ref{fig:A3128substructure}. The peak to the cluster's southwest (A2) saturates our significance maps at $\sigma = 4.42$ at apertures with Schirmer filters larger than $R_S= 3500$ pixels. The northeast peak (A1) does not achieve its maximum significance until $R_S = 4500$ pixels, at which point it, too, saturates our significance maps.  The two substructures merge into one large $4.42~\sigma$ structure in all significance maps made with apertures 6000 pixels and larger. However, the two substructures can be still distinguished in $S/N$ maps up through an aperture size of 9000 pixels. 
Both aperture masses achieve their maximal signal-to-noise at $R_S = 8000$ pixels: the southwest peak with $S/N = 6.6$, and the northeast peak with $S/N = 5.6$. Recalling that the Schirmer filter's weight peaks at $0.10R_S$, the corresponding angular scale of the substructures is 3.5 arcminutes.   The northeast peak (A1) is spatially coincident with the brightest cluster galaxy, and both peaks are near the optical center of the cluster. Note that both of these features are spatially coincident with small-scale features in the ``southwest peak" of X-ray emission (cf. Tables~\ref{CentroidCoords} and~\ref{SubstructureCoords}), and are distinct from the northeastern lobe of X-ray emission which is the high-redshift cluster ACT-CL J0330-5227. 

\begin{figure*}[htb]
\begin{center}
\includegraphics[width=0.60\textwidth]{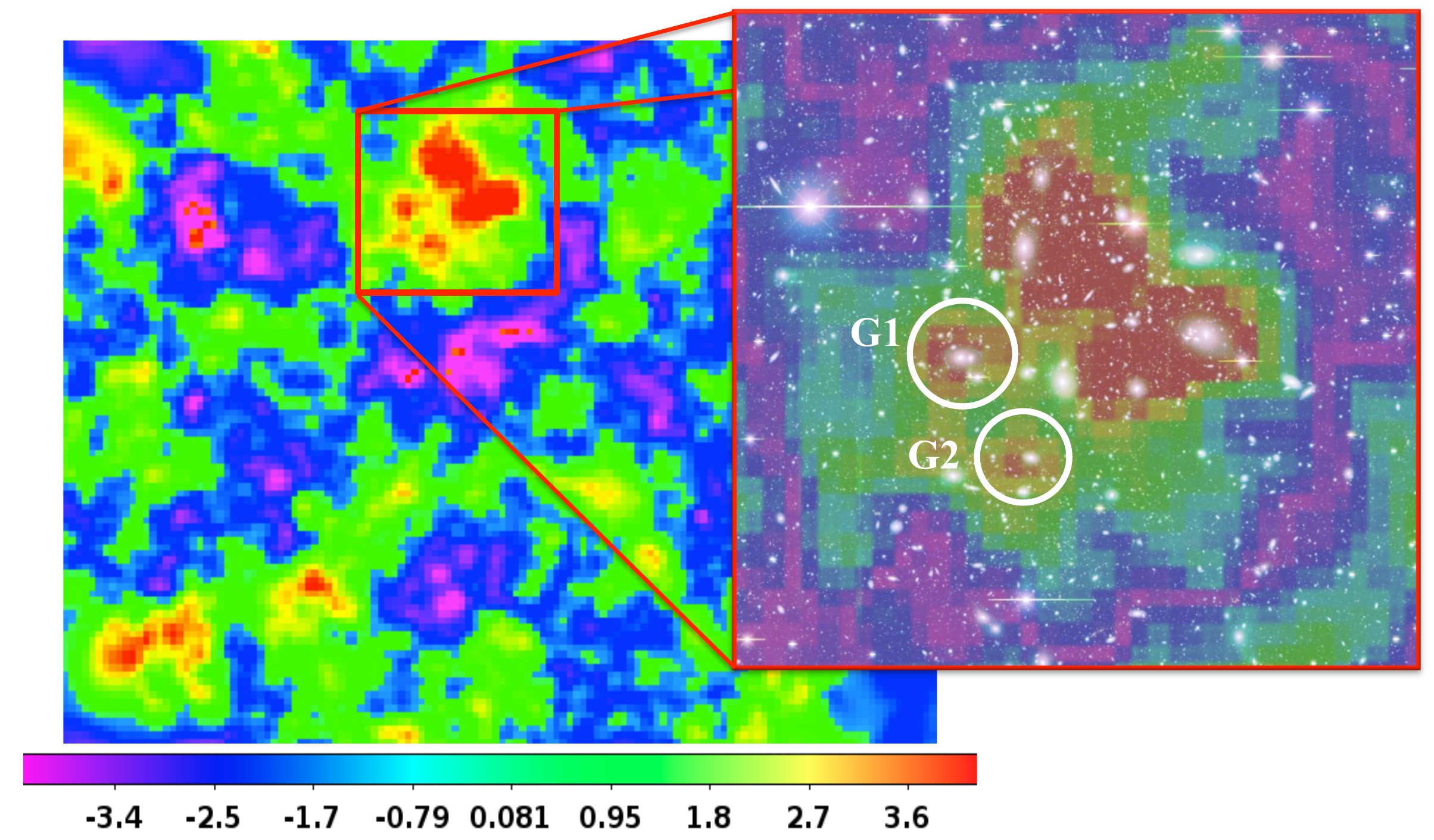}
\caption{\footnotesize{WL significance map made with a Schirmer aperture of $R_S = 5500$ pixels ($= 24'$).  At this filter size size, the two substructures in A3128 have nearly merged, but other structures surrounding the cluster are now visible. {\em Inset:}  Magnified view of the WL significance map, overlaid on a $zrg$ composite image of A3128. The two high significance peaks circled in white are likely associated with galaxy groups recently accreted onto the cluster.}}
\label{fig:InfallingGroups}
\end{center}
\end{figure*}

Within the same range of Schirmer apertures, we detect at high significance two substructures adjacent to the primary A3128 peak, circled in white in Figure~\ref{fig:InfallingGroups} and with coordinates listed in Table~\ref{SubstructureCoords}. At $R_S$ = 5500 pixels (corresponding to an object size of 2.4'), the topmost aperture mass (G1) achieves its peak significance of $4.42\sigma$  while its neighboring feature (G2) achieves $4.17\sigma$ with a 5000 pixel aperture. At their respective characteristic aperture sizes, the features achieve $S/N$ of 5.1 and 4.3. By $R_S$ = 6000, both features have merged into the larger signal of A3128. From their spatial coincidence with knots of galaxies at the same redshift as A3128~(\citealp{2002AJ....123.1216R}), these two peaks are likely associated with the cores of groups that have fallen into the cluster. 
\LongTables
\begin{deluxetable}{ccc}
	\tabletypesize{\scriptsize}
	\tablecaption{\footnotesize{Substructures Identified in Convergence Maps}\label{SubstructureCoords}}
	\tablewidth{0pt}
	\tablehead{
	\colhead{} & \colhead{$\alpha$} & \colhead{$\delta$} \\
	\colhead{ID} & \colhead{(J2000.0)} & \colhead{(J2000.0)}
	}
	\startdata
 			A1/``Northeastern Peak" & $3^h 30^m 02^s.6$&$-52\degree38' 33''$\\
			A2/``Southwestern Peak"  & $3^h 30^m 28^s.4$&$-52\degree33' 57''$ \\
			B1 & $3^h 35^m 32^s.9$ & $-53\degree38' 15''$\\
			B2 & $3^h 34^m 50^s.0$ & $-53\degree35' 40''$\\
			B3 & $3^h 32^m 42^s.3$& $-53\degree29' 38''$ \\
			G1 & $3^h 31^m 15^s.9$& $-52\degree38' 26''$ \\
			G2 & $3^h 29^m 55^s.0$& $-52\degree34' 50''$ \\
	\enddata

\end{deluxetable}%

The three high-$\sigma$ features towards the bottom left of Figure~\ref{fig:A3128substructure}(a) match up to regions of mass enhancement in the signal-to-noise image of Figure~\ref{fig:A3128substructure}(c).  The leftmost peak (B1) is just at the edge of the detector, but peak B2 is less than an arcminute from the rich group ACO S 366 ($z\sim0.06$). We surmise that peak B1 is caused by a combination of detector edge effects and a genuine DM enhancement associated with ACO S 366. Peak B1 achieves its maximum $\sigma = 4.42$ with a Schirmer filter size of 5000 pixels, and peak B2 does the same at $R_S = 4500$ pixels. The corresponding angular scales of these two aperture masses are 2.2' and 2.0', respectively. 
Beginning at peak B3, an arc of WL signal with $S/N\sim2.5-3.0$ stretches across the bottom of Figure~\ref{fig:A3128convergence}(b) and ends at the location of A3125 ($\alpha = 3^h 27.4^m,\delta = -53\degree31'$). While there is a corresponding bridge of galaxies in the DECam image, there are only moderate enhancements in the significance map. We find no significant WL features at the location of A3125 itself.

\subsection{Mass Estimates\label{sec:nfwfits}}

Having constrained the locations of the A3128 primary aperture mass and its substructures with WL convergence maps, we proceed to parametrically fit them with NFW masses following the procedure in \textsec\ref{sec:NFWfitting}. 

\subsubsection{Single NFW fit to Primary A3128 Aperture Mass}
Centering a single NFW peak at the A3128 barycenter yields a best-fit mass of $M_{200} = 10.0 \pm 2.3 \times 10^{14} M_{\odot}$. The corresponding $r_{200}$ for this mass is 2.1 Mpc, which at the distance of A3128 spans about 7000 pixels on the observation. Uncertainty is estimated through a jackknife approach wherein we randomly resample 50\% of the galaxy catalog and recompute the best-fit mass. The variance of 2000 realizations is taken as a measure of the NFW fitting procedure's internal consistency and becomes the error bar on the best-fit mass.

\subsubsection{Simultaneous NFW fits to Abell 3128 and ACT-CL J0330-5227}
NFW shear profiles were fit to the A3128 barycenter and ACT-CL J0330-5227 simultaneously, using both the full galaxy sample and the high-redshift subsample ($z_{gal} > 0.44$) from \textsec\ref{sec:hiz}. Parametric masses resulting from those fits are displayed in the left column of Table~\ref{NFWfits1} under the Centroid 1 heading. We also performed fits with the A3128 shear profile centered on the southwest substructure peak (``Centroid 2," right column of Table~\ref{NFWfits1}).  Errors on these and all other NFW fits in Table~\ref{NFWfits1} are obtained through the variance of a 50\% random resampling of the catalog. All fits were subject to 2000 resamplings except the full galaxy sample/Centroid 1 fits, which were resampled 4000 times. 

 In fits made with the full galaxy sample, the masses of the two clusters sum to about $2.2\times 10^{15} M_{\odot}$, but the allocation of this mass between A3128 and ACT-CL J0330-5227 varies significantly ($\gtrsim 1~\sigma$ error bars) depending on the A3128 center chosen. When the A3128 shear profile is centered at the barycenter, the high-$z$ cluster is assigned a lower mass than when the A3128 profile is centered at the southwest substructure peak.  A similar picture emerges in the 2-peak fits with the high-redshift galaxy sample ($z_{gal} > 0.44$): the combined mass of A3128 and ACT-CL J0330-5227 remains the same (there, $2.7\times 10^{15} M_{\odot}$) regardless of A3128 centroid chosen, but the distribution of this mass between the two clusters varies significantly.  We note that compared to the fits with the full galaxy sample, the variance of the A3128 and ACT-CL J0330-5227 masses in Table~\ref{NFWfits1} are slightly higher -- an expected by-product of the smaller sample size. 
 
Table~\ref{NFWfits1} also contains the results of a two-peak fit made with the intermediate-redshift subsample of galaxies ($0.16 < z_{gal} < 0.4$) of \textsec\ref{sec:hiz}.  As expected when using galaxies exclusively in front of the high-redshift cluster, no mass was allocated to ACT-CL J0330-5227. The mass assigned to A3128 in this round of two-peak fits 
underestimates the other values in Table~\ref{NFWfits1} at the $\gtrsim 1.5~\sigma$ level. Because of the low source density of the intermediate-redshift subsample (7-10 arcmin$^{-2}$), and moreover because those galaxies have an unfavorable $D_S/D_{LS}$, the underestimate likely reflects a WL signal too weak to beat down the random shape noise of background galaxies.
\LongTables
\begin{deluxetable*}{ccccc}
	\tabletypesize{\scriptsize}
	\tablecaption{\footnotesize{NFW Masses for High-Significance Aperture Masses}\label{NFWfits1}}
	\tablewidth{0pt}
		\tablehead{			
  			\colhead{}& \multicolumn{2}{c}{A3128 Centroid 1} & \multicolumn{2}{c}{A3128 Centroid 2} \\
  			\hline
			\colhead{Galaxy Sample} & \colhead{A3128 Mass\tablenotemark{a}} & \colhead{High-$z$ Mass} & \colhead{A3128 Mass\tablenotemark{b}} & \colhead{High-$z$ Mass}\\
 			\colhead{}&\colhead{($10^{14} M_{\odot}$)} & \colhead{($10^{14} M_{\odot}$)} & \colhead{($10^{14} M_{\odot}$)} & \colhead{($10^{14} M_{\odot}$)}
		}
		\startdata
 			Full $z_{gal}$ sample & 11.5 $\pm$ 1.7 & 10.3 $\pm$ 3.2 & 7.3 $\pm$ 1.3 & 15.3 $\pm$ 2.4\\
			$z_{gal} > 0.44 $& 13.8 $\pm$ 2.4 & 14 $\pm$ 4.0 & 8.0 $\pm$ 2.1 &  19.5 $\pm$ 3.5\\
			$0.16 < z_{gal} < 0.4$& 7.0 $\pm$ 1.9 & $<$ 0.4\tablenotemark{c} & 3.5 $\pm$ 1.3 & $<$ 0.4\tnote{c}\\
		\enddata
           \begin{centering}
           \tablenotetext{a}{Centered at cluster barycenter: $\alpha = 3^h 30^m 16^s.5,\delta = -52\degree33' 57''$}
           \tablenotetext{b} {Centered at SW substructure peak: $\alpha = 3^h 30^m 02^s.6,\delta = -52\degree38' 33''$}
	\tablenotetext{c}{Fits with this galaxy sample consistently assigned to the high-redshift cluster minimum mass in the allowed range, down to $1\times 10^{13} M_{\odot}$} 
	\end{centering}
\end{deluxetable*}%

\begin{figure*}[tb]
 	\centering
	\includegraphics[width=0.85\textwidth]{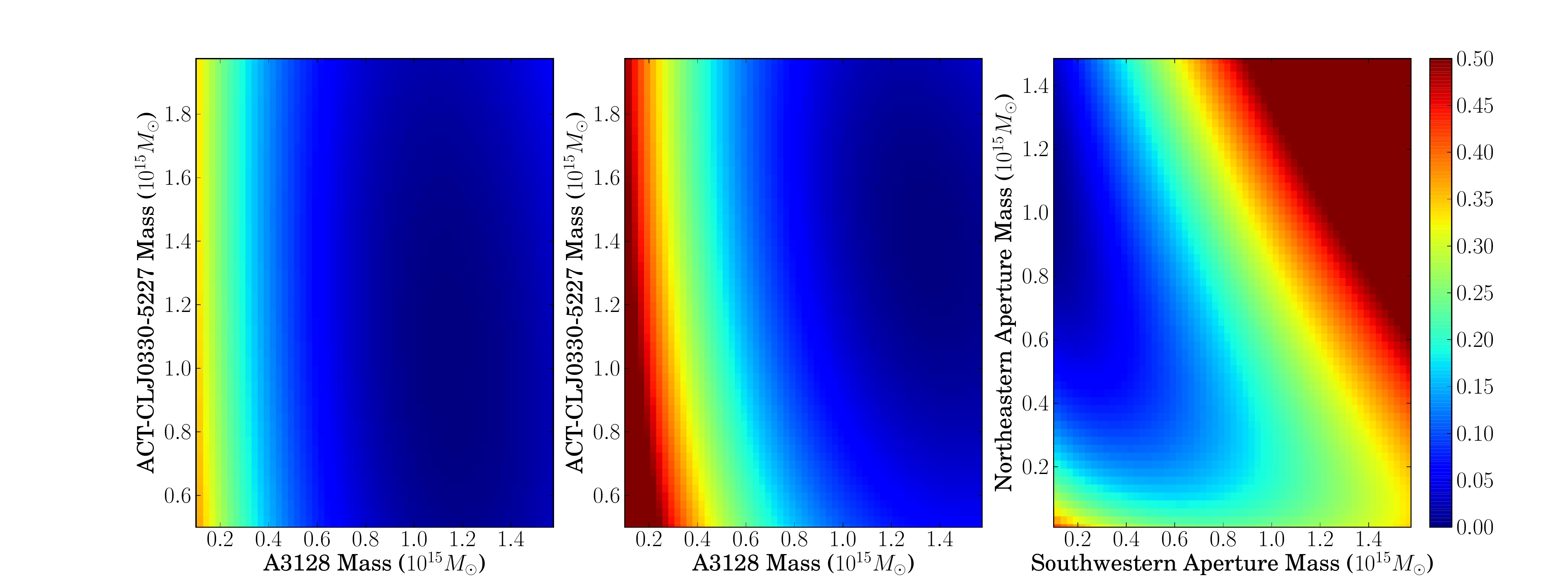}
	\caption{\footnotesize{$\chi^2$ landscapes for fits of NFW shear profiles to the WL peaks identified in the reconstructions. For ease of viewing, the values have been rescaled to each distribution's respective $(\chi^2 - \chi^2_{min})\times 1000$. {\em Left:} Residuals from parametric mass fits to ACT-CL J0330-5227 and A3128 (centered at its barycenter) made with the full galaxy sample. {\em Center:} Residuals from parametric mass fits to ACT-CL J0330-5227 and A3128 (centered at its barycenter) with $z_{gal} > 0.44$. {\em Right:} Residuals from fits to the two A3128 substructures with the mass of ACT-CL J0330-5227 fixed to $1.0\times10^{15} M_{\odot}$.}}
	\label{fig:NFWchi2}
\end{figure*}

The $\chi^2$ landscape shown in Figure \ref{fig:NFWchi2}(a) indicates that the A3128 mass is well constrained by the full galaxy sample fits, but that they only place marginal constraints on the high-redshift cluster mass. Eliminating all galaxies in front of the high-redshift cluster leads to slightly better constraints on the mass of ACT-CL J0330-5227, as evidenced by the tighter $\chi^2$ ellipse about the best-fit masses in Figure~\ref{fig:NFWchi2}(b). Coupled with the fact that the high-$z$ cluster mass experiences significantly greater variance in all fits, Figure \ref{fig:NFWchi2} suggest that A3128 dominates the WL signal embedded in the background galaxy catalogs.  
 

\subsubsection{NFW Fits to Substructures}

We also attempted to fit masses to the two high-significance, small-aperture substructures identified within the A3128 peak (A1 and A2 in Table~\ref{SubstructureCoords}). The lowest $\chi^2$ fit assigned a mass of $1.02 \times 10^{15} M_{\odot}$ to the northeastern substructure and the minimum boundary value mass of $0.5 \times 10^{14} M_{\odot}$ to the southwestern. These results were replicated in a simultaneous 3 NFW shear profile fit which also included ACT-CL J0330-5227. Although the combined masses of the two substructures matches the best-fit A3128 mass in Table~\ref{NFWfits1}, their allocation appears inconsistent with the larger size and higher $S/N$ of the southwestern substructure and the fact that it saturates significance maps sooner than the northeastern substructure. The shunting of the A3128 mass to the northeast aperture mass likely reflects its proximity to the cluster barycenter (400 pixels away) compared to the southwest aperture mass (over 1000 pixels away). Figure~\ref{fig:NFWchi2}(c) supports an interpretation where the total mass of the cluster is constrained to $1.0\times10^{15} M_{\odot}$ by this set of fits, but that the two substructures cannot be resolved with individual NFW profiles.  

Finally, NFW masses were fit to the two infalling groups identified in Figure~\ref{fig:InfallingGroups} (G1 and G2 in Table~\ref{SubstructureCoords}) in a simultaneous 4 NFW shear profile fit with A3128 and ACT-CL J0330-5227. To guarantee the distinctness of G1 and G2 from the shear profiles of the two clusters, A3128 and ACT-CL J0330-5227 masses were fixed to their respective full sample/Centroid 1 values of $1.1\times10^{15} M_{\odot}$ and $1.03\times10^{15} M_{\odot}$. The fits returned masses of $2.7\pm 3.2\times10^{13} M_{\odot}$ for G1, and $2.2\pm1.7\times10^{13} M_{\odot}$ for G2, where the error bars come from the variance of 2000 jackknife resamplings, as before. 

\subsubsection{Global Tests of NFW Fitting Procedure}

\begin{figure}[tb]
 	\centering
	\includegraphics[width=0.5\textwidth]{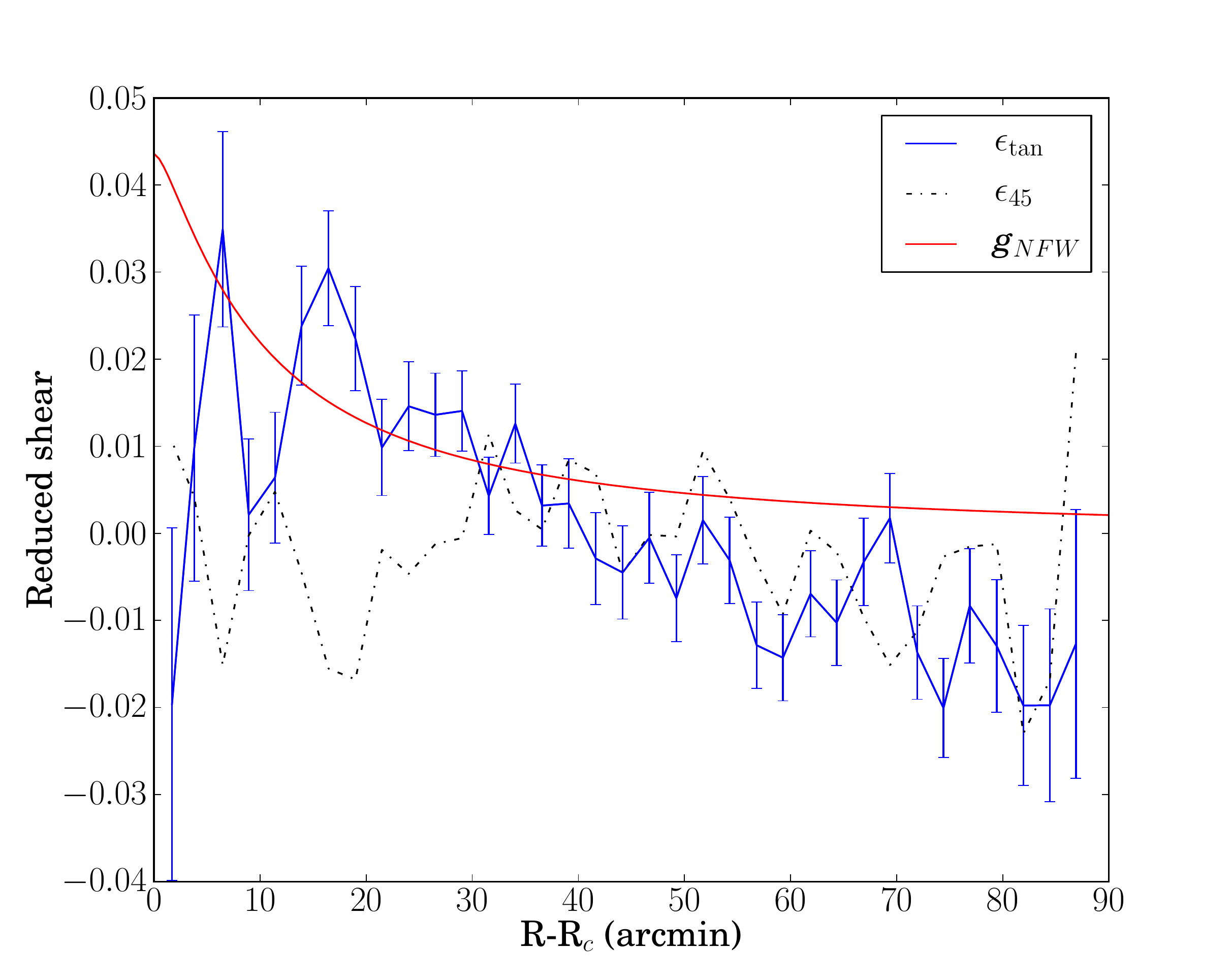}
	\caption{\footnotesize{ Tangential shear profile for the 1-peak NFW fit to the A3128 barycenter (solid red line), overplotted on the azimuthually averaged tangential ellipticity signal of background galaxies (solid blue line). The dashed line is the B-mode signal of galaxy ellipticity. Error bars on the galaxy ellipticity signal are the value of reduced shear in a radial bin ($\sim \langle e_{\tan}\rangle/\sqrt{2}$) divided by $\sqrt{N}$ galaxies in that bin. The best-fit mass of $1.1\times10^{15} M_{\odot}$ was used for the theoretical curve.}}
	\label{fig:shearprofile}
\end{figure}

 In Figure~\ref{fig:shearprofile}, the best-fit A3128 single-peak tangential shear profile is compared with the azimuthally averaged galaxy ellipticity signal. We again emphasize that mass estimates in this work do not result from fitting 1-D NFW profiles to binned galaxy ellipticities; the shear profiles shown below are for comparison only.
The wide area of the A3128 observation and large number of background galaxies allows for a fine radial binning and detailed inspection of galaxy ellipticity signal. The negative value in the first radial bin is a  manifestation of the PSF undercorrection near the cluster center discussed in \textsec\ref{sec:OverallDetection}. At large $R-R_c$, the galaxy ellipticity signal should approach zero; however Figure~\ref{fig:shearprofile} shows a noticeable downward trend in galaxy ellipticity. As in \textsec\ref{sec:OverallDetection}, we attribute this negative galaxy signal to an undercorrection of the PSF ``pincushion," manifesting as a net tendency of galaxies at large distances from the cluster center to be aligned radially outwards. To avoid underestimating the WL signal in A3128 when computing the masses shown in Table~\ref{NFWfits1}, we truncated the catalog to galaxies 58 arcminutes from the cluster center. Compared to NFW fits using an untruncated galaxy sample, this had the effect of raising the masses by 10\%.

\begin{figure*}[htb]
 	\centering
	\includegraphics[height=0.34\textwidth]{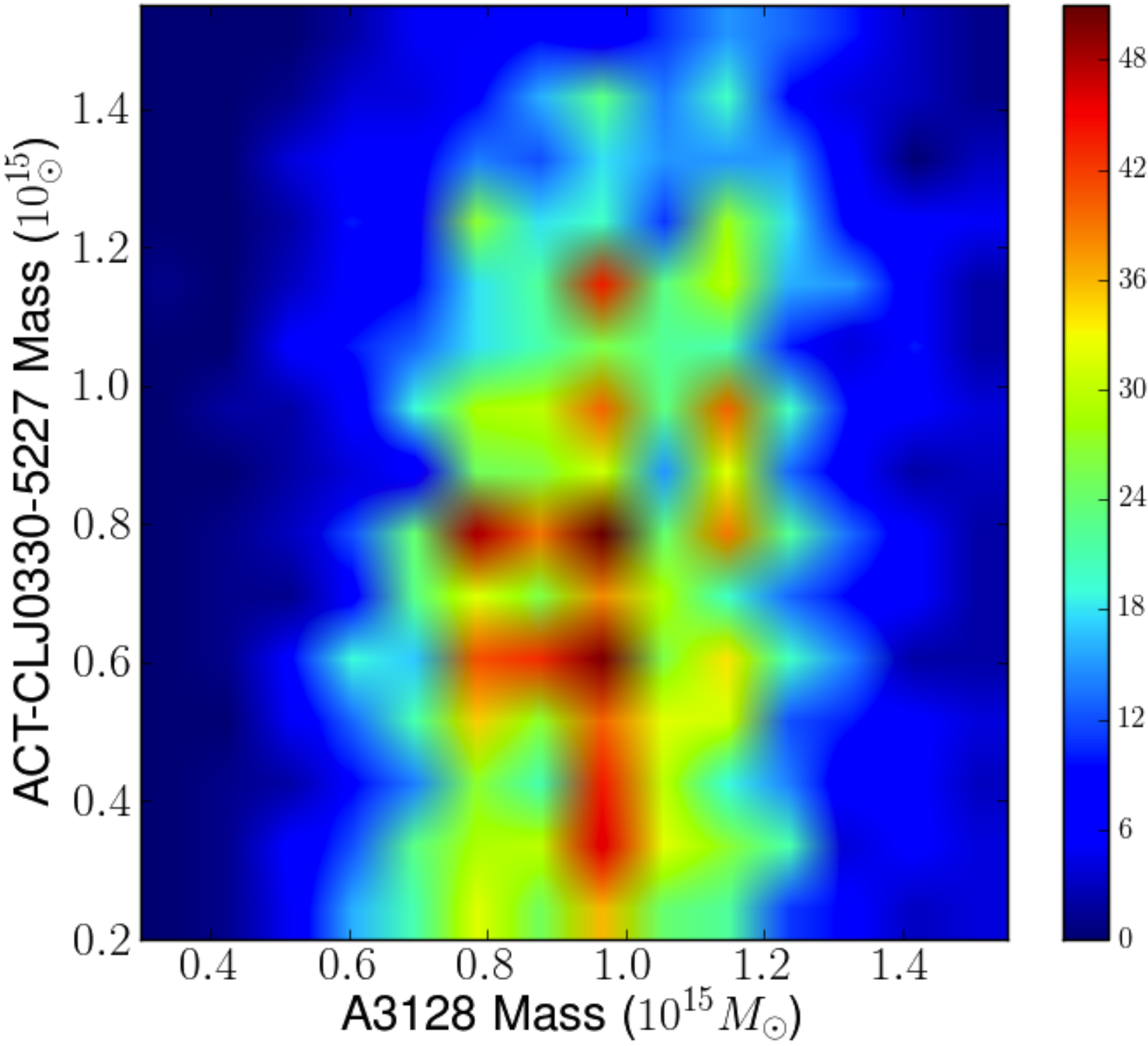}
	\includegraphics[height=0.345\textwidth]{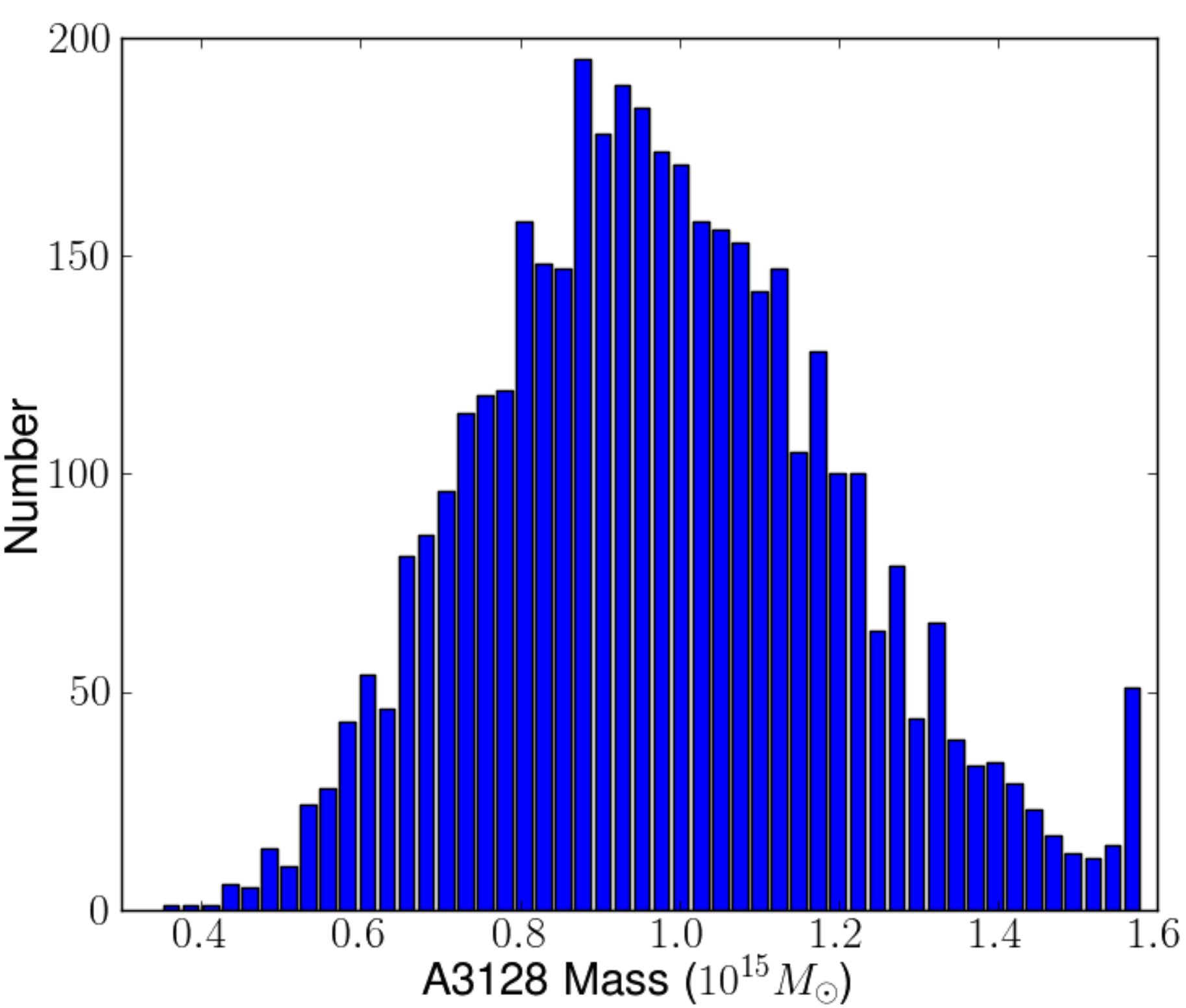}
	\caption{\footnotesize{{\em Left:} 2-D histogram of parametric masses for A3128 (with NFW shear profile centered at the barycenter) and ACT-CL J0330-5227 returned by 4000 jackknife resamplings of the full (untruncated) galaxy catalog. {\em Right:} Distribution of most probable NFW masses for A3128 with an unconstrained ACT-CL J0330-5227 mass. This plot is equivalent to projecting the 2-D histogram at left along its y-axis of ACT-CL J0330-5227 masses. The spike at the end of the x-axis is caused by projecting all masses greater than $1.55\times10^{15}$ into the highest bin.}}
	\label{fig:occupancyNFW}
\end{figure*}
Uncertainties on cluster masses in Table~\ref{NFWfits1} presume that our measurements obey Gaussian statistics. This assumption may be tested using the distribution of cluster masses returned by the jackknife procedure. The left panel of Figure~\ref{fig:occupancyNFW} shows a 2-D histogram of cluster masses returned by 4000 random resamplings of the full galaxy catalog. The right panel of Figure~\ref{fig:occupancyNFW} shows the distribution of allowed A3128 NFW masses returned in the resampling, i.e., the left panel collapsed along the y-direction of ACT-CL J0330-5227 mass. Starting at the the median mass ($\sim 1.0\times10^{15} M_{\odot}$), we sum 34.1\%, 47.7\% and 49.9\% of the returned masses on either side of the distribution. The equivalent 68\% confidence interval is $(7.6,12.2)\times10^{14} M_{\odot}$ and the 95\% confidence interval is $(5.9,15.6)\times10^{14} M_{\odot}$.  When summing over 99\% of the returned mass range, the bound on the low-mass end is $4.6 \times10^{14} M_{\odot}$. However, the upper boundary value of the sampled mass range is reached before we can find an equivalent upper bound. The 68\% confidence interval is symmetric about the median (i.e., $1.0_{-2.4}^{+2.2}\times10^{15} M_{\odot}$), and is also roughly equivalent to the 1 $\sigma$ variance of Table~\ref{NFWfits1}. However, the 95\% and 99\% confidence intervals are not equivalent to 2 and 3 $\sigma$, nor are they symmetric about the median. This skewness to high masses suggests that the distribution of allowed masses departs from Gaussianity at the endpoints, though it is Gaussian near the best-fit value. 

 The finite resolution of our significance maps leads to to an uncertainty in the coordinates of WL peak centroids.  Any resulting mis-centering of NFW profiles might bias the reported masses. This potential systematic was probed through a slew of mass fits in which the identified centers of A3128 and ACT-CL J0330-5227 were individually shifted north, south, east, and west by 200 pixels (the size of the centroid uncertainty). The mass of ACT-CL J0330-522 varied an average of 7.3\% from its Table~\ref{NFWfits1} value, while the mass of A3128 only varied by 2.7\%. Since the variations are smaller than the error bars reported in Table~\ref{NFWfits1}, the uncertainty of centroid coordinates is probably not a dominant source of error in our results.

%

\begin{figure*}
	\centering
	\includegraphics[width=0.90 \textwidth]{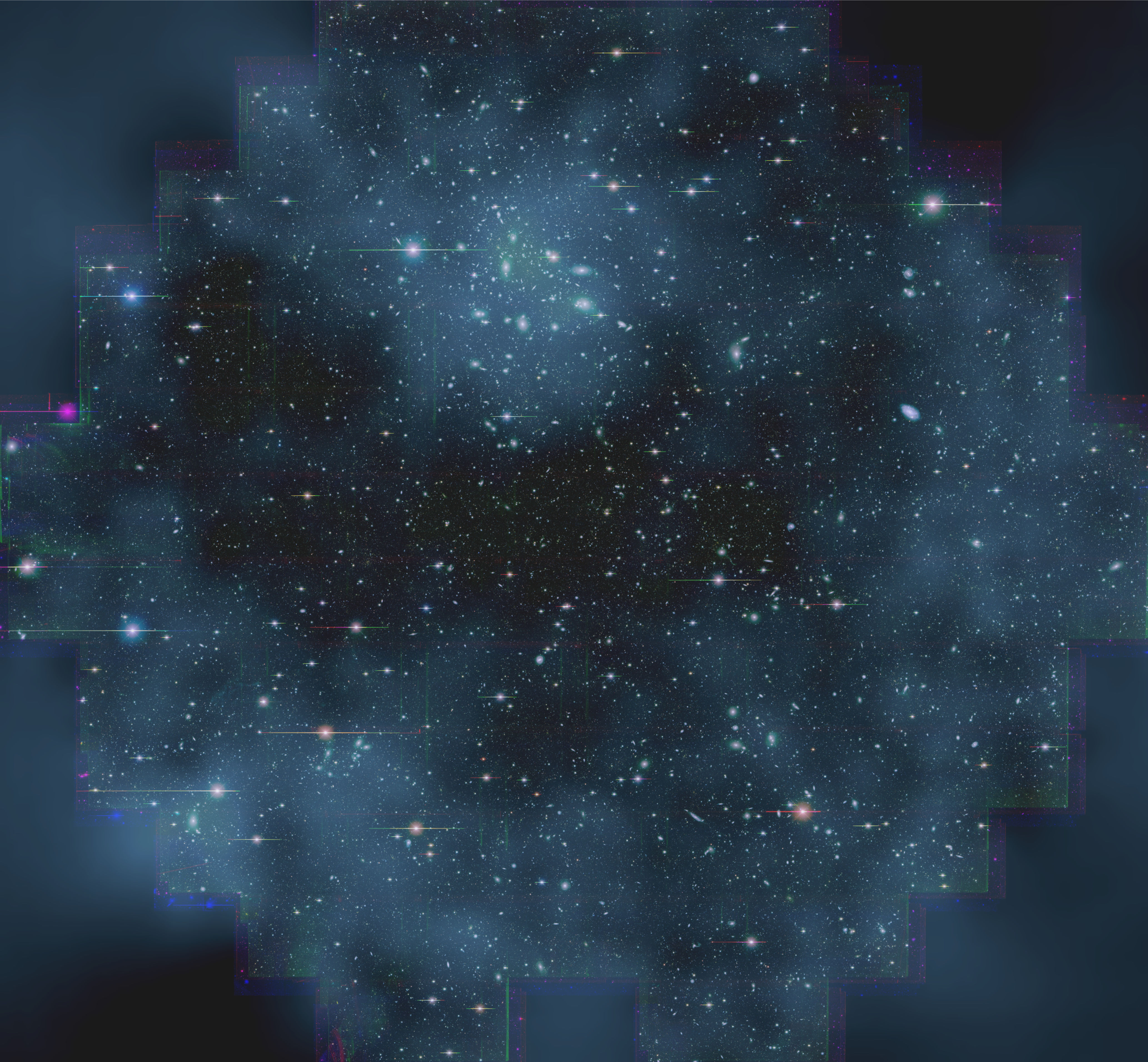}
	\caption{\footnotesize{Abell 3128 raw convergence map for a 10000 pixel Schimer aperture size, colorized and superimposed on $zrg$ composite image. The bright clump on the top left is Abell 3128; the fainter clumps on the bottom left are from infalling groups at $z = 0.06-0.08$. The dispersed arc of signal along the bottom is from Abell 3125, a cluster disturbed by a recent passage near Abell 3128 (Werner \etal 2007).}}
	\label{fig:PrettyConvergence}
\end{figure*}

\section{DISCUSSION\label{sec:Discussion}}

With increasing $R_S$, the smaller aperture mass peaks detected in the periphery of A3128 such as  ACO S 366 (B1 in Figure \ref{fig:A3128substructure}) display the gradual increase and decrease of significance and $S/N$ expected from \textsec\ref{sec:Convergence}. However, the primary cluster saturates significance maps with $\sigma = 4.42$ at all Schirmer radii considered. In addition, the two A3128 substructures merge into the primary cluster signal at $R_S = 6000$ in significance maps, 
but remain distinct in $S/N$ maps through much larger kernel sizes. These results suggest that the primary cluster and its substructures are detected at much higher confidence than $4.42~\sigma$. They also underscore a fundamental limitation of the significance maps, which are only as good as the number of random noise iterations performed. Unfortunately, the gain in significance is a slow function of the number of random maps. For example, to achieve $\sigma = 4.89$ requires $1\times10^6$ iterations, and $\sigma = 5.33$ requires $1\times10^7$ iterations or 100 times more computation time. Given unlimited computational resources, the two substructures could be teased apart from the primary cluster signal at $R_S > 5000$ pixels and firmly assigned an equivalent $\sigma$ of detection confidence.

We note that the departure from Gaussianity in Figure~\ref{fig:occupancyNFW} does not invalidate our use of significance maps to evaluate WL signal. The ``$\sigma$" of a significance map is a Gaussian-equivalent confidence, representing a signal pixel's distance from the mean, and is rooted in an exact pixel probability distribution (see \textsec\ref{sec:Convergence}). 
Should the distribution of random noise pixels have the same skew positive seen in Figure~\ref{fig:occupancyNFW}, the $\sigma$ of our significance maps -- as well as our $S/N$ values, which assume a Gaussian noise distribution -- would in fact only be pseudo-Gaussian. However, the {\it relative} pixel-to-pixel enhancements in significance maps would still be genuine: a 2 $\sigma$ peak is still 95.4\% less likely than the mean value of WL signal, a 3 $\sigma$ peak is still 99.7\% less likely than the mean value of WL signal, etc. For this reason, the significance maps and the distribution of allowed NFW masses are complementary: while jackknife resampling makes no assumptions about the distribution of WL signal pixels, it must be remade for each aperture mass as it contains no spatial information. On the other hand, the significance maps (whether strictly Gaussian or not) do reveal the 2-D distribution of projected mass. 

While ACT-CL J0330-5227, A3128, and the A3128 substructures are all unambiguously detected in WL reconstructions at very high significance, their parametric mass fits are subject to some degeneracy. Both in simultaneous NFW fits to the two clusters and in simultaneous fits to the A3128 substructures, the sum of the masses is more tightly constrained than their respective magnitudes. This is evident from all three $\chi^2$ plots of Figure~\ref{fig:NFWchi2}, in which the best-fit ellipses tilt about a line of constant mass.  Table~\ref{NFWfits1} also attests to the degeneracy of mass fits: with both the full and high-redshift galaxy samples, the aggregate mass of ACT-CL J0330-5227 and A3128 is conserved, but their respective magnitudes vary at the $\lesssim 1~\sigma$ level depending on the chosen A3128 NFW profile centroid. The significant dependence of cluster mass on A3128 centroid signals that the exact assignment of mass in our two-peak fits is driven by sources near the center of the cluster, and that galaxies further away only bear the overall signal. As it happens, the source density in our observation is depressed near the cluster center -- a consequence of studying low redshift clusters whose member galaxies subtend wide angles. This explains the degeneracy in our parametric mass fits, and particularly our inability to resolve the A3128 substructures with NFW shear profiles.

 Figure~\ref{fig:shearprofile} shows a downward trend in galaxy ellipticity at distances past 50-60 arcminutes from the cluster center, which we attributed above to an under-correction of the ``pincushion" in the DECam PSF. A residual radial gradient in the PSF was also invoked explain the non-zero correlation at small scales visible in Figure~\ref{fig:correlfuncs}, and the negative galaxy signal in Figure~\ref{fig:PSFstarSN}. To avoid underestimating the WL signal in the observation, the NFW mass fits were supplied exclusively with galaxies within 58 arcminutes of the cluster center.  In future work, we will attempt to improve our PSF circularization scheme, perhaps adopting the method of \cite{2013MNRAS.429.3627M} to reject exposures with unsatisfactory PSF correction. 

Aside from highlighting the shortcomings of our PSF correction scheme, Figure~\ref{fig:shearprofile} has another interesting feature. The large number of galaxies in our observation allows for a fine binning of the galaxy ellipticity signal which, in turn, allows us to appraise the success of an NFW profile in describing it. Ignoring the inner 5 arcminutes, it is clear that the galaxy ellipticity distribution tends to zero faster than predicted by an NFW fit: galaxy ellipticities are already consistent with zero by 35 arcminutes, while the NFW fit does not settle to zero on the scale of the image. Cosmological simulations such as have been published by \cite{2014arXiv1407.4730D} report a similar conclusion, that the density profiles of cluster-size halos fall off more quickly than predicted by either the NFW or Einasto mass models. In future work, we will attempt to fit the galaxy ellipticity distributions with the latest generation of mass distributions. 

\section{CONCLUSIONS\label{sec:Conclusion}}

The significance maps and the distribution of most probable masses in Figure~\ref{fig:occupancyNFW} are completely independent ways of confirming the WL signal from Abell 3128. Significance maps employ the aperture mass statistic of Equation~\ref{aperturemassstat} to sum up background galaxies' tangential ellipticity and return the convergence signal at each point in the observation. Figure~\ref{fig:occupancyNFW} results from the parametric fitting of an NFW halo shear profile to the tangential ellipticities of background galaxies. With either method, the weak lensing signal from Abell 3128 is detected at high significance. In particular, the probability distribution of A3128 mass obtained by randomly resampling of the full galaxy catalog skews to high masses, from which it follows that low values of A3128 mass are more strongly disfavored than higher masses. Our confidence in the WL detection of A3128 is thus higher than might be indicated by significance maps alone.

Given a sufficient density of background galaxies (and with the caveat that the number of degrees of freedom decreases with added NFW peaks), the tools we have developed allow for the fitting of NFW masses to an arbitrary number of substructures. Our average source density of 20 galaxies per arcminute was enough to constrain the mass of Abell 3128 to $(1.1\pm 0.2) \times 10^{15} M_{\odot}$ through both the simultaneous NFW fits with the high-redshift cluster ACT-CL J0330-5227 and the substructure mass fits. However, the background galaxy density drops to  $\sim$ 13 arcmin$^{-2}$ near the cluster center because of the large apparent size of its member galaxies. Combined with the unfavorable distance ratio of background galaxies in our observations, these issues prevent the fitting of NFW masses to the two central substructures. Moving forward, we expect that all WL studies of low-redshift galaxy clusters will be similarly affected, and this places a lower limit on the source density required to resolve their substructures with NFW shear profiles.  

Since they can be mistaken for mass substructures, interloping high-redshift clusters like ACT-CL J0330-5227 pose a problem for systematic studies of low-redshift clusters. Even when they are not fully resolved in WL convergence maps, high-redshift background clusters can have knock-on effects on NFW mass estimates because their presence may blur the location of the lower-redshift cluster's barycenter. The tests performed in \textsec\ref{sec:hiz} were a means of vetting the extraction of photometric redshifts from deep DECam imaging, as we expected to see WL signal from ACT-CL J0330-5227 only in galaxies with $z \ge 0.44$.  The success of the redshift tests suggests a natural means of authenticating potential mass substructures: a localized WL signal enhancement that appears only when distant galaxies are used -- like ACT-CL J0330-5227 in Figure~\ref{fig:hiz2} -- is almost certainly an interloping high-redshift cluster. 


Abell 3128 is one of the lowest redshift clusters to have been studied with weak gravitational lensing in such detail. The advent of wide-angle cameras such as DECam makes the systematic studies of low-redshift clusters possible. In particular, the work on A3128 presented here is the pilot for our WL study of the mass distributions of a complete, volume-limited sample of massive galaxy clusters between $0.04<z<0.1$ using DECam. We hope that this work will in turn become the low-redshift anchor to a systematic observational measurement of the evolution of mass substructure in clusters of galaxies. The first of its kind, such a study would deepen our understanding of how the first galaxies assembled themselves into clusters. In addition, the presence of substructure may induce biases in the mass determination of clusters that must be precisely calibrated to extract precision cosmological information from the cluster mass function and its evolution. Upcoming surveys such as the Dark Energy Survey (DES) and LSST will measure averaged lensing signals from thousands of clusters to calibrate the cluster mass function. Therefore, empirically measuring the role of substructure in mass measurements is now of great importance. 


\acknowledgments
 
The authors thank Dara Norman for her efforts in performing all observations made during the DECam science verification phase, including those of Abell 3128 used in this study.  The authors also thank the anonymous referee for suggesting many useful improvements to the analysis. This project used data obtained with the Dark Energy Camera (DECam), which was constructed by the Dark Energy Survey (DES) collaborating institutions: Argonne National Lab, University of California Santa Cruz, University of Cambridge, Centro de Investigaciones Energeticas, Medioambientales y Tecnologicas-Madrid, University of Chicago, University College London, DES-Brazil consortium, University of Edinburgh, ETH-Zurich, University of Illinois at Urbana-Champaign, Institut de Ciencies de l'Espai, Institut de Fisica d'Altes Energies, Lawrence Berkeley National Lab, Ludwig-Maximilians Universitat, University of Michigan, National Optical Astronomy Observatory, University of Nottingham, Ohio State University, University of Pennsylvania, University of Portsmouth, SLAC National Lab, Stanford University, University of Sussex, and Texas A\&M University. Funding for DES, including DECam, has been provided by the U.S. Department of Energy, National Science Foundation, Ministry of Education and Science (Spain), Science and Technology Facilities Council (UK), Higher Education Funding Council (England), National Center for Supercomputing Applications, Kavli Institute for Cosmological Physics, Financiadora de Estudos e Projetos, Funda‹o Carlos Chagas Filho de Amparo a Pesquisa, Conselho Nacional de Desenvolvimento Cient'fico e Tecnol—gico and the MinistŽrio da Cincia e Tecnologia (Brazil), the German Research Foundation-sponsored cluster of excellence ``Origin and Structure of the Universe" and the DES collaborating institutions.

\end{document}